\documentclass[11pt]{article}
\usepackage{charter} 
\usepackage{fullpage}
\usepackage{titlesec}
\usepackage[hyphens]{url}     
\usepackage[hidelinks]{hyperref}       
\usepackage{booktabs}       
\usepackage{amsfonts}       
\usepackage{amssymb}        
\usepackage{mathrsfs}
\usepackage{nicefrac}       
\usepackage{microtype}      
\usepackage{bm}				
\usepackage{cite}			
\usepackage{graphicx}		
\usepackage{mathtools}		
\usepackage{algorithmic}	
\usepackage{algorithm}
\usepackage{amsthm}			
\usepackage{enumitem}		
\usepackage{multirow}       
\usepackage{tabularx}	    
\usepackage{hhline}
\usepackage{arydshln}		
\usepackage{xcolor}
\usepackage{siunitx}
\usepackage{diagbox}

\definecolor{lightgreen}{rgb}{.9,1,.9}

\newcolumntype{L}[1]{>{\raggedright\arraybackslash}m{#1}}
\newcolumntype{C}[1]{>{\centering\arraybackslash}m{#1}}
\newcolumntype{R}[1]{>{\raggedleft\arraybackslash}m{#1}}

\theoremstyle{plain} 

\def\defn{\,\coloneqq\,}
\def\argmin{\mathop{\mathrm{arg\,min}}} 

\def\lim{\mathop{\mathsf{lim}}} 

\def\max{\mathop{\mathsf{max}}}
\def\mse{\mathop{\mathsf{mse}}}

\def\log{\mathsf{log}}

\def\proposed{DeCAF}

\def\Bbm{{\bm{B}}}
\def\cbm{{\bm{c}}}

\def\ebm{{\bm{e}}}

\def\xbm{{\bm{x}}}

\def\ybm{{\bm{y}}}
\def\ybmhat{\widehat{\bm{y}}}

\def\nbm{{\bm{n}}}
\def\ubm{{\bm{u}}}

\def\vbm{{\bm{v}}}

\def\Abm{{\bm{A}}}

\def\Hbm{{\bm{H}}}
\def\Ibm{{\bm{I}}}
\def\Rbm{{\bm{R}}}

\def\xbmhat{{\widehat{\bm{x}}}}

\def\Tsf{{\mathsf{T}}}

\def\Dsf{{\mathsf{D}}}

\def\Isf{{\mathsf{I}}}

\def\Rsf{{\mathsf{R}}}
\def\Usf{{\mathsf{U}}}

\def\C{\mathbb{C}}
\def\R{\mathbb{R}}

\def\Mcal{{\mathcal{M}}}
\def\Ncal{{\mathcal{N}}}

\def\Lcal{{\mathcal{L}}}
\def\Fcal{{\mathcal{F}}}

\def\Rcal{{\mathcal{R}}}

\def\sin{{\textit{sin}}}
\def\cos{{\textit{cos}}}

\def\epsilonbm{\bm{\epsilon}}

\newlength{\tempdima}
\newcommand{\rowname}[1]
{\rotatebox{90}{\makebox[\tempdima][c]{\textbf{#1}}}}
\usepackage{caption}
\usepackage{subcaption}

\makeatletter
\renewcommand{\maketitle}{\bgroup\setlength{\parindent}{0pt}
\begin{flushleft}
  \textbf{\huge\@title}
  
  \vspace{2em}
  
  {\Large\@author}
\end{flushleft}\egroup
}
\makeatother

\begin{document}

\title{Recovery of Continuous 3D Refractive Index Maps from Discrete Intensity-Only Measurements using Neural Fields}

\author{
{\LARGE Renhao Liu$^{\dagger,1}$},\\ 
\medskip
{\LARGE Yu Sun$^{\dagger,1}$},\\
\medskip
{\LARGE Jiabei Zhu$^{3}$}\\
\medskip
{\LARGE Lei Tian$^{3,4}$}, and\\
\medskip
{\LARGE Ulugbek~S.~Kamilov$^{\ast,1,2}$}\\
\vspace{1em}
\emph{$^1$ Department of Computer Science and Engineering,~Washington University in St.~Louis, MO 63130, USA}\\
\emph{$^2$ Department of Electrical and Systems Engineering,~Washington University in St.~Louis, MO 63130, USA}\\
\emph{$^3$ Department of Electrical and Computer Engineering,~Boston University, MA 02215, USA}\\
\emph{$^4$ Department of Biomedical Engineering,~Boston University, MA 02215, USA}\\
\vspace{1em}
$^{\ast}$ \emph{Correspondence author}: \texttt{kamilov@wustl.edu}\\
$^{\dagger}$ \emph{These authors have contributed equally to the work and are ordered alphabetically} \\
\vspace{1em}
\emph{Renhao Liu conducted the work at Washington University in St. Louis and is now at Google Inc.} \\
\emph{Yu Sun conducted the work at Washington University in St. Louis and is now at Caltech}
}

\date{}

\maketitle

\newpage

\section*{Abstract}
Intensity diffraction tomography (IDT) refers to a class of optical microscopy techniques for imaging the 3D refractive index (RI) distribution of a sample from a set of 2D intensity-only measurements. The reconstruction of artifact-free RI maps is a fundamental challenge in IDT due to the loss of phase information and the missing cone problem. Neural fields (NF) has recently emerged as a new deep learning (DL) approach for learning continuous representations of physical fields. NF uses a coordinate-based neural network to represent the field by mapping the spatial coordinates to the corresponding physical quantities, in our case the complex-valued refractive index values. We present \proposed~as the first NF-based IDT method that can learn a high-quality continuous representation of a RI volume from its intensity-only and limited-angle measurements. The representation in \proposed~is learned directly from the measurements of the test sample by using the IDT forward model, without any ground-truth RI maps. We qualitatively and quantitatively evaluate DeCAF on the simulated and experimental biological samples. Our results show that \proposed~can generate high-contrast and artifact-free RI maps and lead to up to $2.1\times$ reduction in MSE over existing methods.

\section*{Main Text}
Refractive index (RI) measures optical density that determines the interaction between light and matter within a sample.
The real part of RI characterizes the phase while its imaginary part characterizes the absorption. RI can thus serve as an endogenous source of optical contrast for imaging samples without staining or labeling.
By quantitatively characterizing the three-dimensional (3D) distribution of the RI, one can visualize cellular or subcellular structures useful for morphogenesis~\cite{KimK.etal2016}, oncology~\cite{Yamada.etal2007}, cellular pathophysiology~\cite{Kim.etal2018}, biochemistry~\cite{Cooper.etal2013}, and beyond (see the review papers~\cite{Park.etal2018,Jin.etal2017}).

\begin{figure}[t!]
\begin{center}
\includegraphics[width=\linewidth]{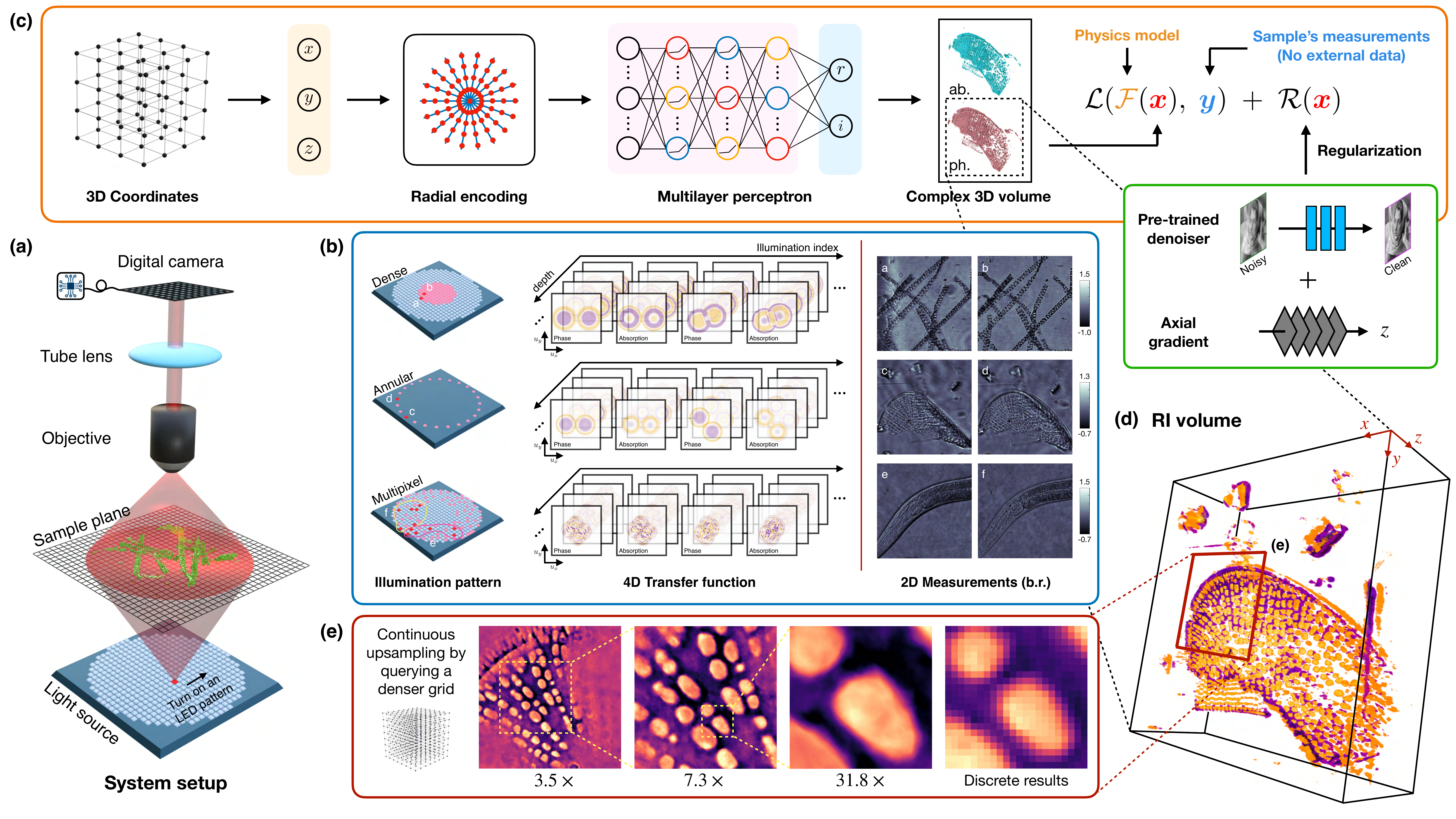}
\end{center}
\caption{
\textbf{Conceptual illustration of \proposed~for IDT.}
\textbf{(a)} \proposed~reconstructs the RI volume by learning a neural field parameterized by a multilayer perceptron (MLP). The network is trained to map the 3D coordinate $(x,y,z)$ to the corresponding RI value by minimizing a loss that penalizes measurement mismatch and imposes regularization.
\textbf{(b)} Our IDT system uses a programmable LED array to illuminate a sample from different angles and uses a digital camera to record the intensity measurements of the scattered light. By changing the illumination patterns, our system can implement different IDT modalities. 
\textbf{(c)} Our experiments consider three illumination patterns, that is, \emph{dense}, \emph{annular}, and \emph{multiplexed} illuminations, each of which corresponds to a different formulation of the forward model. 
\textbf{(d)} \proposed~can reconstruct high-quality 3D RI maps from intensity-only and limited-angle measurements. \textbf{(e)} \proposed~learns a continuous representation and can render samples on a pixel grid of the desired density (illustration with $3.5\times$, $7.5\times$, and $31.8\times$ upsampling).
}
\label{Fig:scheme}
\end{figure}

\emph{Intensity diffraction tomography (IDT)} is a recent technique for recovering 3D RI maps of a sample by measuring the light it scatters. In the standard IDT setup, a sample is illuminated multiple times from different angles, and a set of two-dimensional (2D) intensity projections are captured by the camera~(see Figure \ref{Fig:scheme}(a)). A tomographic image reconstruction algorithm is then used to computationally reconstruct the desired 3D RI distribution from the set of 2D measurements. Unlike traditional optical diffraction tomography (ODT) that uses interferometry to record the complex-valued light fields~\cite{Park.etal2008,Sung.etal2009,Kamilov.etal2015b}, IDT only measures the squared amplitude of the scattered light, leading to an easy setup on standard transmission optical microscopes with inexpensive hardware modifications. Such flexibility has spurred different IDT variants integrating object scanning~\cite{Gbur.etal02, Jenkins.etal2015}, angled illumination~\cite{Tian.Waller2015, Chen.etal2016, Ling.etal18, LiJ.etal2018}, pupil engineering~\cite{Wang.etal2011, Nguyen.etal2017}, and multiple scattering~\cite{Chowdhury.etal2019, Chen.etal2020}. Setups achieving high resolution~\cite{Chowdhury.etal2019} and fast acquisition~\cite{LiJ.etal2019} have also been reported.

Despite the rich literature on IDT, image reconstruction remains a fundamental challenge. The first issue is that the phase of the scattered light field is missing from the measurements, resulting in a \emph{nonlinear} measurement system that is not characterizable by the classical linear Fourier diffraction theory~\cite{Kak.Slaney1988}.
This rules out the usage of the standard filtered-backprojection methods and calls for advanced computational algorithms. The second issue is the well-known \emph{missing cone} problem, causing elongation of the reconstructed  object along the optical axis ($z$ dimension) and hence reduction of the axial resolution.
The missing cone problem is due to the limited-angle tomographic setup, where illuminations can come only from one side of the sample plane with a limited range for angle variation (less than  $\sim40^\circ$ in our setups). This leads to an incomplete coverage of the 3D Fourier spectra with a cone-shape missing region in the axial direction.
These missing phase and missing cone problems make image reconstruction in IDT a severely ill-posed \emph{inverse problem}. 

Regularization methods are widely-used for mitigating ill-posed nature of many inverse problems. These methods are based on minimizing a cost function consisting of a data-fidelity term and a regularization term, where the former uses a physical model to quantify the mismatch between the predicted and acquired measurements, while the latter promotes solutions that are consistent with {\it a priori} knowledge of the sample.
For example, the least-squares loss and Tikhonov regularizer ($\ell_2$-penalty) are widely-used for obtaining closed-form solutions to inverse problems~\cite{Ling.etal18}. The work on \emph{plug-and-play priors} has generalized the notion of image priors to implicit regularizers characterized by image denoisers~\cite{Venkatakrishnan.etal2013,Sreehari.etal2016,Chan.etal2016,Ahmad.etal2020}. Recently, \emph{deep learning (DL)} has emerged as a powerful framework for image reconstruction. A traditional DL reconstruction is based on training a \emph{convolutional neural network (CNN)} over a large dataset to learn a mapping from low-quality images to their high-quality counterparts. The state-of-the-art performance of such methods has been demonstrated in X-ray computed tomography~\cite{Kang.etal2017,DJin.etal2017}, magnetic resonance imaging~\cite{Zhu.etal2018,Aggarwal.etal2019}, optical tomography~\cite{Sun.etal2018,Li.etal2018}, and seismic imaging~\cite{ZhangZ.etal2020} (see the reviews~\cite{Wang.etal2020,Liang.etal2020,Adler.eta2021}).
While DL has significantly improved image reconstruction in many modalities, traditional DL methods are impractical for image reconstruction in IDT where it is difficult to acquire high-quality ground-truth RI maps in experiments. Although a physics-based simulator has been proposed to generate datasets for training IDT artifact-suppressing CNNs, the results are still limited by the mismatch between the simulation and experiments~\cite{Alex.etal2021}.

\emph{Neural fields (NF)} is a recent DL framework that has gained popularity in computer vision and graphics for representing and rendering 3D scenes using coordinate-based deep neural networks~\cite{Sitzmann.etal2019a,Sitzmann.etal2019}. It is worth mentioning that while NF was suggested to be the most appropriate term~\cite{Hinton2021, Piala.Clarck2021}, this idea currently goes by various names in the vision/graphics literature, including \emph{neural coordinate-based representations} or \emph{neural implicit models}. It has been shown that NF can learn a high-quality representation of a complex scene from a sparse set of data without any external training dataset. Motivated by this property, we propose \emph{\textbf{De}ep \textbf{C}ontinuous \textbf{A}rtifact-free RI \textbf{F}ield (\proposed)} as a first NF-based IDT method for learning a high-quality continuous 3D RI map from intensity-only and limited-angle measurements without any external training dataset of ground-truth RI maps. Figure~\ref{Fig:scheme} provides a conceptual illustration of \proposed. The key features of \proposed~are as follows:
\begin{itemize}

\item The central component of \proposed~is a \emph{multilayer perceptron (MLP)}, which is a fully-connected (non-convolutional) deep network, for learning a function that maps 3D coordinates $(x,y,z)$  to the corresponding complex-valued RI values. The trained MLP thus provides a \emph{continuous} neural representation of the RI map. The RI value at \emph{any} spatial location can be retrieved by querying the trained MLP with the corresponding coordinate. 
By decoupling representation from an explicit voxel grid, \proposed~can efficiently store large 3D volumes.

\item \proposed~is a \emph{self-supervised} method, meaning that it does not require training using an external dataset of ground-truth RI maps. This is possible since the same MLP is used at every 3D location, enabling it to learn natural redundancies and correlations within a RI volume. The MLP is trained directly at test time by using only the IDT measurements of the sample that we seek to reconstruct. The IDT forward model is used to map the output of MLP to the intensity measurements and using the gradient back-propagation to update the MLP weights.

\item \proposed~enables easy integration of additional prior knowledge on the unknown sample using an explicit regularization term in the loss function. In this paper, we explored the potential of such synergistic integration by including an anisotropic 3D regularizer that separately imposes penalties in the $x$-$y$ plane and $z$ direction. 
Specifically, the $x$-$y$ penalty uses a deep denoising CNN pre-trained on \emph{natural images} to remove additive white Gaussian noise (AWGN)~\cite{Sun.etal2020, Sun.etal2021}, and the $z$ penalty is based on one-dimensional (1D) total variation. 
While our denoising CNN was not trained explicitly on RI images, we show through ablation studies that it improves the performance by mitigating noise and imaging artifacts.
\end{itemize}

The pipeline of the proposed method is visually illustrated in Figure~\ref{Fig:scheme}(c). In the training phase, the input of \proposed~is a set of spatial coordinates, $\cbm=\{(x_i,y_i,z_i)\}_{i=1}^n$, taken from a pre-defined grid.
\proposed~first maps the input coordinates to encodings using a non-trainable radial expansion, followed by a standard fully connected neural network to map the encodings to the RI values at the input coordinates. We introduced a novel type of encoding called \emph{radial encoding}, which facilitates high-reconstruction, artifact-free reconstruction of RI maps (see details in Method). 
\proposed~is trained to solve the following optimization with an objective consisting of a measurement loss $\Lcal$ and regularizer $\Rcal$
\begin{align}
\label{Eq:Loss}
&\phi^\ast = \argmin_\phi\{\Lcal(\Fcal(\xbm), \ybm) + \mathcal{R}(\xbm)\} \nonumber\\
&\text{such that}\quad\xbm = \mathcal{M}_\phi(\cbm),
\end{align}
where $\xbm$ is the predicted RI map, $\ybm$ is the intensity measurements of the test sample, $\Fcal$ is the IDT forward model, and $\Mcal_\phi$ is the MLP (which includes the radial encoding) parameterized by weights $\phi$. 
Note that the test measurements are the only input required in  \proposed.
After the optimal $\phi^\ast$ is learned, one can render the test sample on a voxel grid with arbitrary density by simply querying $\Mcal_{\phi^\ast}$ using the corresponding coordinates, as illustrated in Figure~\ref{Fig:scheme}(e).

Prior applications of NF include novel view synthesis~\cite{Mildenhall.etal2020,Martin.etal2020,Yu.etal2020}, dynamic scene representation~\cite{Park.etal2020,Peng.etal2021,Li.etal2021}, object lightning~\cite{Srinivasan.etal2020,Wizadwongsa.etal2021}, and computed tomography~\cite{Reed.etal2021}. 
Our work has several novel contributions to the existing NF literature: \emph{(i)} \proposed~is the first method that considers learning NF by accounting for the diffraction and scattering effects due to the wave nature of the light, while the existing work in the area has focused on ray-tracing models in graphics. \emph{(ii)} \proposed~is the first NF-based method that considers the recovery of the phase information from \emph{intensity-only} data.  \emph{(iii)} \proposed~is the first method that combines an \emph{implicit} MLP regularization with an additional \emph{explicit} image regularizer (for example, based on a deep denoiser) to achieve the best of both worlds, that is to improve over the separate usage of an implicit and explicit regularization. \emph{(iv)} \proposed~introduces \emph{radial encoding} as a novel type of encoding layer for improving the ability of NF to represent complex samples. The details of the network architecture and the learning procedure of \proposed~are provided in the Methods and Supplement. 
In the next section, we present both qualitative and quantitative results showing the ability of \proposed~to reconstruct high-quality RI maps.

\subsection*{Results}
\label{Sec:Results}

\subsubsection*{Experimental Validation}

We validated the ability of \proposed~on experimentally collected IDT data  to recover high-quality RI maps with accurate biological features and minimal artifacts. We used \proposed~on four biological samples, including \emph{spirogyra}, \emph{diatom}, \emph{human buccal epithelial cells}, and \emph{Caenorhabditis elegans}. We adopted the existing light-propagation models to formulate the inverse problems associated with the dense~\cite{Ling.etal18}, annular~\cite{LiJ.etal2019}, and multiplexed~\cite{Matlock.etal2019} illumination patterns. 
Since absorption provides lower contrast for the considered samples, we focus on comparing the phase images.
In the subsequent sections, we use $x$, $y$, and $z$ to denote length, width, and depth, respectively.

\begin{figure}[t!]
\begin{center}
\includegraphics[width=\linewidth]{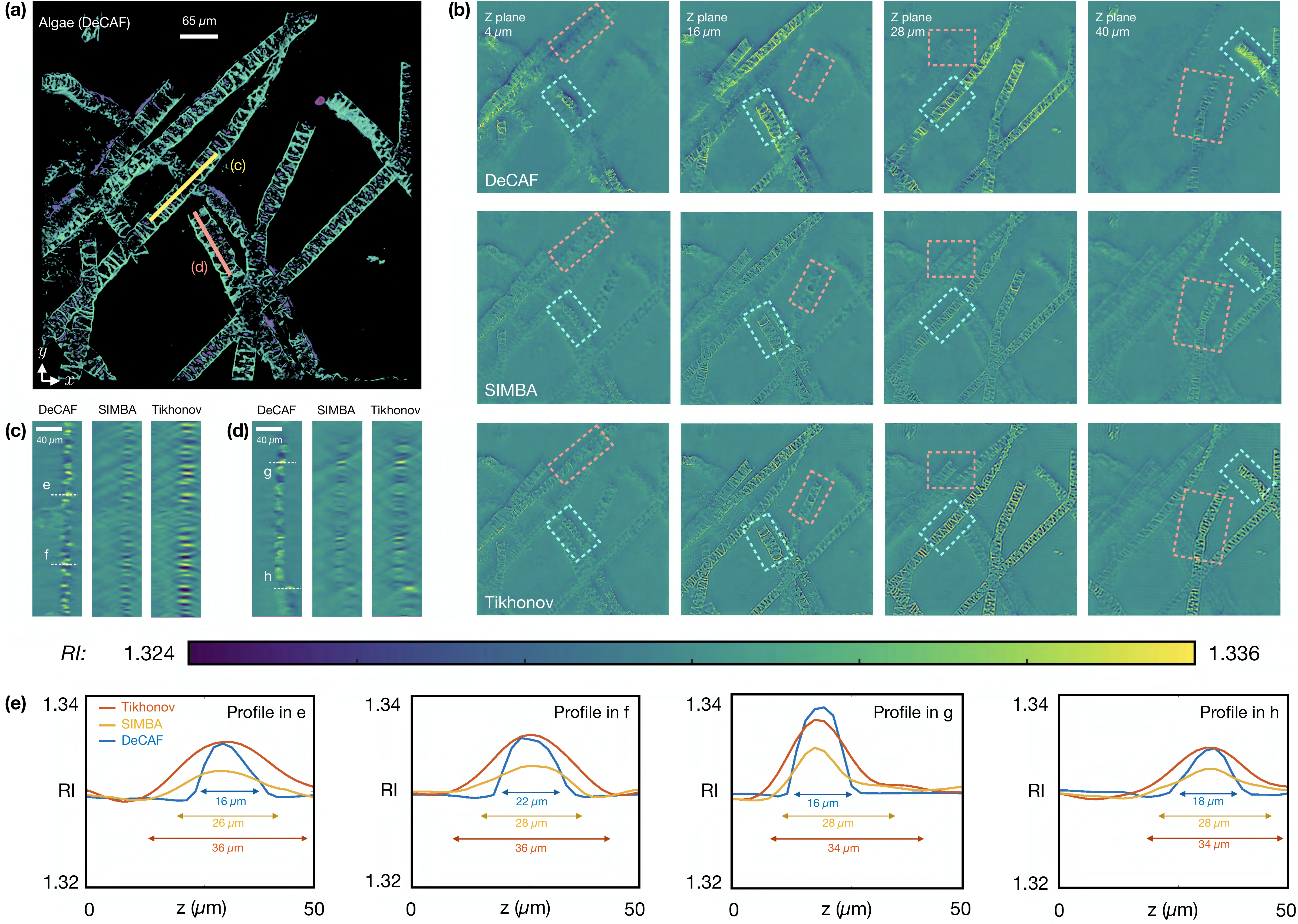}
\end{center}
\caption{
\textbf{Reconstruction of \emph{Spirogyra Algae} acquired by dIDT.} \textbf{(a)} 2D rendering obtained by accumulating all the $z$ slices from \proposed. Scale bar 65 $\mu$m.
\textbf{(b)} Visual comparison of the axial views at $z\in\{4, 16, 28, 40\}\;\mu$ms reconstructed using three methods: \proposed, SIMBA, and Tikhonov. 
\textbf{(c) \& (d)} Axial views corresponding to the colored lines in (a). Scale bar 40 $\mu$m.
\textbf{(e)} Line profiles from the dashed lines in (c) and (d). The label at the upper right of each plot indicates the corresponding dashed line. 
This figure illustrates the ability of \proposed~to reconstruct high-contrast RI maps by also significantly reducing the missing-cone artifacts. Note the quantitative demonstration of the reduction of elongation highlighted in (e).
Additional examples are shown in Supplementary Videos \emph{spirogyra-decaf.mov}, \emph{spirogyra-simba.mov}, and \emph{spirogyra-tikhonov.mov}.
} 
\label{Fig:spirogyra}
\end{figure}

We first show the effectiveness of \proposed~for \emph{dense IDT (dIDT)} on stained spirogyra (Fisher Scientific S68786, embedded in water $\text{RI}\approx1.33$), which is unicellular algae containing helical arrangement of chloroplasts oriented in the 3D space.
We collected in-total $89$ brightfield intensity measurements using a $0.25$ NA objective lens. Two example measurements  are presented in  Figure~\ref{Fig:scheme}(b) (see images a-b).
In the experiment, we compared \proposed~against two existing IDT reconstruction baselines: Tikhonov regularization (\emph{Tikhonov})~\cite{Ling.etal18} and SIMBA~\cite{Wu.etal2020}, as both methods have been extensively validated under similar imaging settings.
SIMBA is a recently proposed model-based algorithm that leverages a deep learning denoiser as an image prior.
The final reconstructed RI volume by each method consists of $40$ axial slices of $1024\times1024$ pixels equally spaced between -$30\,\mu$m and $50\,\mu$m, forming a volume of $665.6\times665.6\times80\,\mu$m$^3$.
Throughout the paper, we define $z=0\,\mu$m as the focal plane.

Figure~\ref{Fig:spirogyra} visualizes the experimental results.
To demonstrate the overall structure of the sample, a rendered 2D image accumulating all $z$ layers of the \proposed~reconstruction is presented in Figure~\ref{Fig:spirogyra}(a). 
As shown, \proposed~successfully reconstructed the spiral structure of the spirogyra.
Figure~\ref{Fig:spirogyra}(b) compares the 2D axial slices obtained by \proposed, SIMBA, and Tikhonov at the depths $z\in\{4, 16, 28, 40\}\,\mu\text{m}$.
The results from \proposed~are visualized in the first row, and those from SIMBA and Tikhonov regularization are visualized in the second and third rows, respectively. 
The results show that \proposed~provides superior axial sectioning ability---meaning that a pattern emerges only in the slices it belongs to and fades rapidly as we move axially to different depths---than the other two methods. This is highlighted by the clarity and sharpness of the spirals (in the cyan box) that appear at a specific depth, showing that \proposed~removes the artifacts (in the white box) generated by the diffraction from the adjacent slices. These artifacts remain in the reconstructions by SIMBA and Tikhonov.
We further evaluate the axial resolution of each reconstruction by comparing the lateral views corresponding to the cutlines shown in Figure~\ref{Fig:spirogyra}(c) and~\ref{Fig:spirogyra}(d). \proposed~significantly reduces the elongation artifacts caused by the missing-cone problem. 
Line profiles presented in Figure~\ref{Fig:spirogyra}(e) quantitatively characterize the reduction of $z$-elongation by \proposed.

\begin{figure}[t!]
\begin{center}
\includegraphics[width=\linewidth]{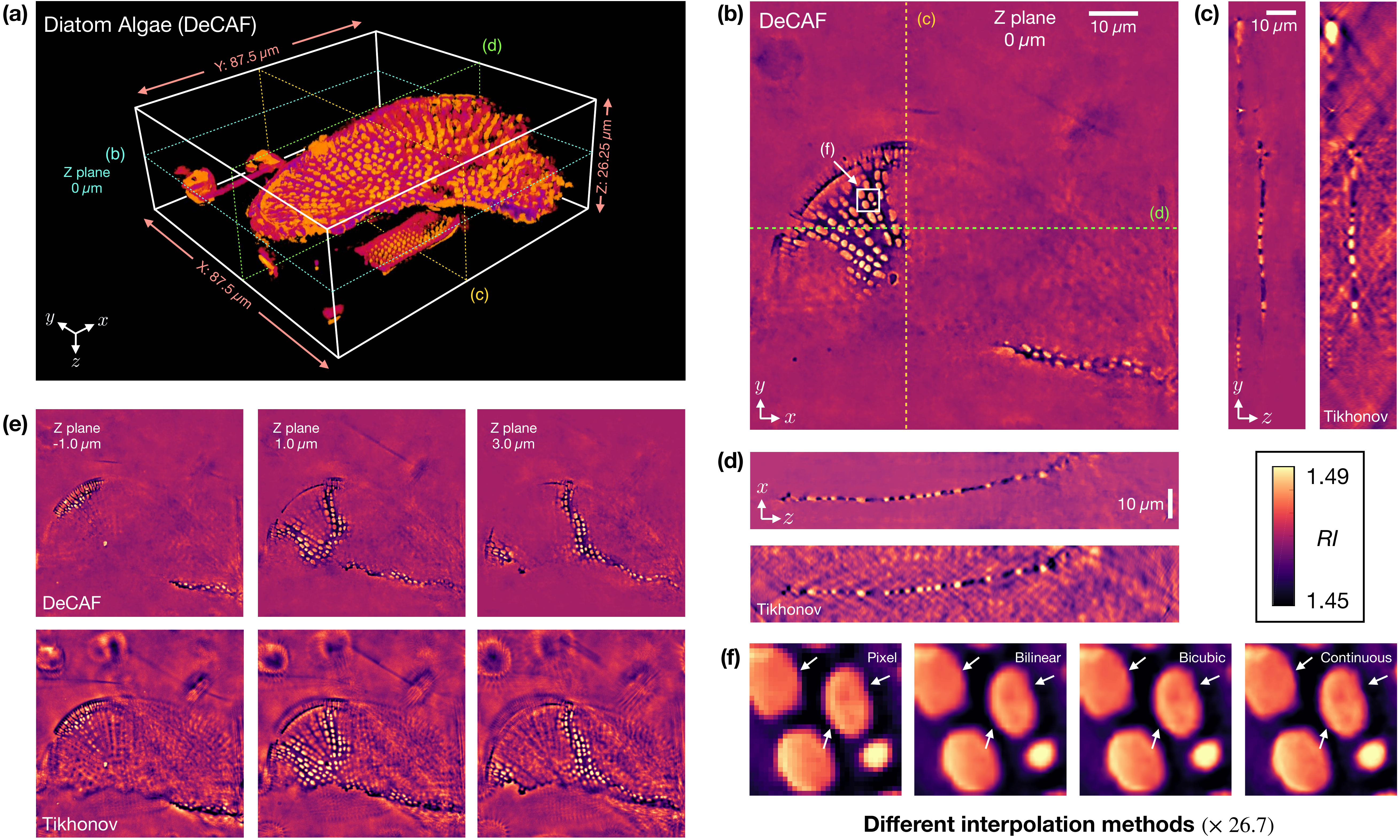}
\end{center}
\caption{
\textbf{Reconstruction of \emph{Diatom Algae} acquired by aIDT.} 
\textbf{(a)} 3D illustration of the \proposed~reconstruction showing the overall structure of the  sample.
\textbf{(b)} Axial view at $z=0\;\mu$m (focal plane) reconstructed by \proposed. Scale bar 10 $\mu$m.
\textbf{(c) \& (d)} $y$-$z$ and $x$-$z$ lateral views corresponding to the colored paths in (b). The results of Tikhonov are also presented for reference. Scale bar 10 $\mu$m. 
\textbf{(e)} Visual illustrations of the axial views at $z\in\{-1.0, 16, 28, 40\}\;\mu$m, highlighting better removal of artifacts compared to Tikhonov.
\textbf{(f)} Visual demonstration of \proposed's ability to perform continuous RI upsampling along the $x$ and $y$ dimensions. \proposed's results are consistent with that of the classic interpolation methods but provide finer details highlighted by the arrows.
Additional examples are shown in Supplementary Videos \emph{diatom-decaf.mov} and \emph{diatom-tikhonov.mov}.
}
\label{Fig:diatom}
\end{figure}

\begin{figure}[t!]
\begin{center}
\includegraphics[width=\linewidth]{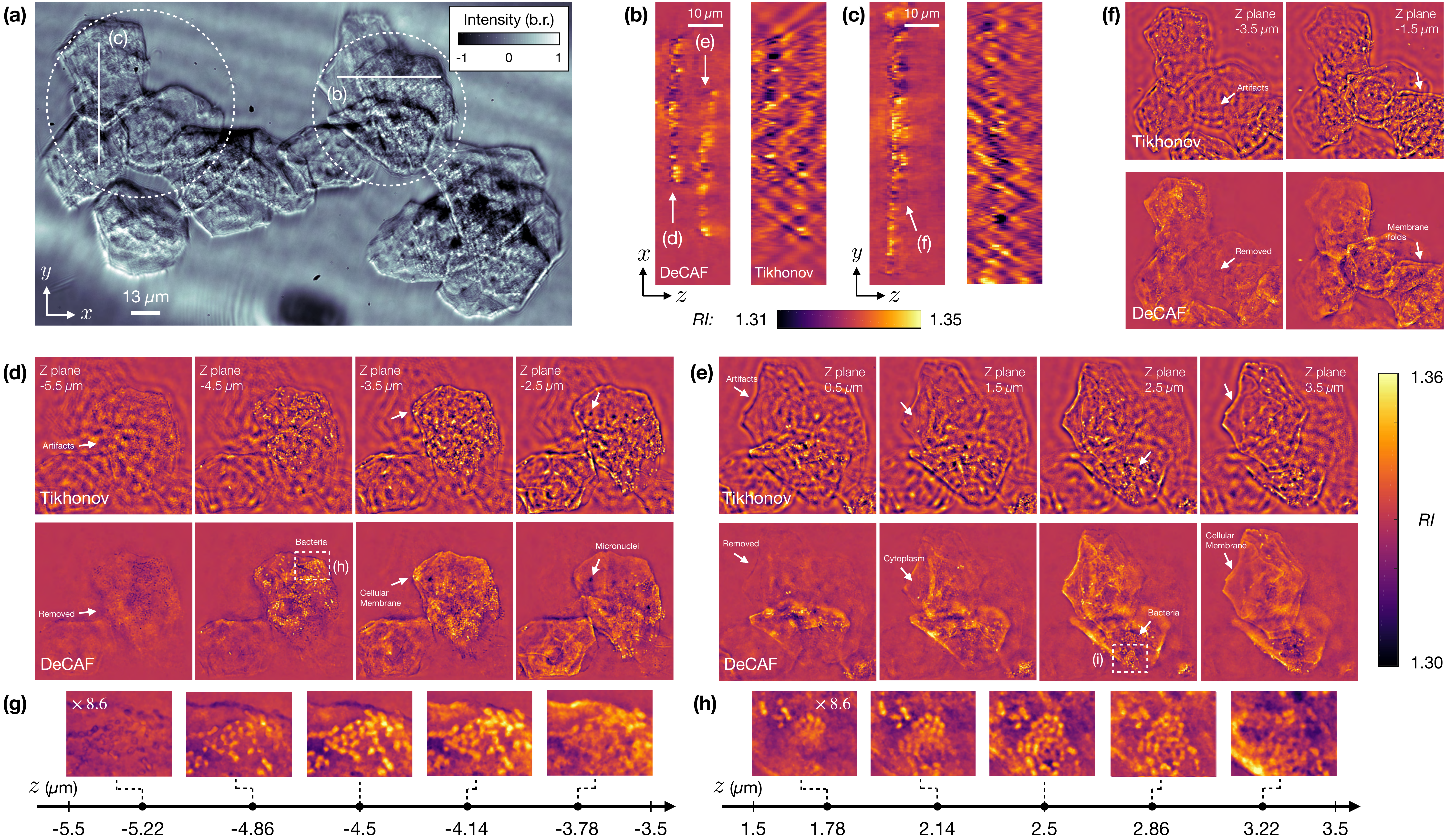}
\end{center}
\caption{
\textbf{Reconstruction of \emph{Human Buccal Epithelial Cell Cluster} acquired by aIDT.} 
\textbf{(a)} Example intensity measurement collected by our aIDT setup for the cell cluster. Note that the background light is removed from the image. Scale bar 13 $\mu$m
\textbf{(b) \& (c)} $x$-$z$ and $y$-$z$ lateral views of the \proposed reconstruction associated with the paths in (a). Superior performance in artifact removal and axial separation is demonstrated over the Tikhonov regularization. Scale bar 10 $\mu$m.
\textbf{(d), (e), \& (f)} The axial views at various depth of the two sub-cell clusters shown in (b) and (c). These results further highlight the strong axial sectioning effects as well as the fine details preserved by \proposed.
\textbf{(g) \& (h)} Visual demonstration of \proposed's ability to perform continuous RI upsampling along the $z$ dimension. 
Note that $\{-5.5, -4.5, -3.5\}\,\mu$m and $\{1.5, 2.5, 3.5\}\,\mu$m are the only axial points used during training.
Smooth and consistent transition in the appearance of bacteria is observed. 
Additional examples are shown in Supplementary Videos \emph{cell-decaf-b.mov}, \emph{cell-decaf-c.mov}, \emph{cell-tikhonov-b.mov}, and \emph{cell-tikhonov-c.mov}.
}
\label{Fig:cells}
\end{figure}

We next applied \proposed~to \emph{annular IDT (aIDT)} to explore its capability for efficient data processing. We imaged two distinct classes of biological samples, including diatom microalgae (S68786, Fisher Scientific) and unstained human epithelial buccal cells.
The former is a unicellular algae with regular arrangement of punctae, while the latter is a complex cell environment consisting of intracellular bacteria.
We acquired $24$ intensity images using a $0.65$~NA objective lens under oblique illuminations for each sample.
The diatom and cell cluster samples are fixed in glycerin gelatin ($\text{RI}\approx1.47$) and water, respectively.
We used Tikhonov as the baseline method for comparison. 

Figure~\ref{Fig:diatom} presents the results for diatom algae. Two example measurements  are provided in Figure~\ref{Fig:scheme}(b) (see images c-d).
Both \proposed~and Tikhonov were configured to reconstruct $52$ slices of $700\times700$ pixels equally spaced between -$10\mu$m to $16\mu$m, forming a volume of $113.75\times113.75\times26\,\mu$m$^3$. 
The 3D illustration of the volume reconstructed by \proposed~is presented in Figure~\ref{Fig:diatom}(a), demonstrating the overall reconstruction quality.
Figure~\ref{Fig:diatom}(e) presents the slices reconstructed by each method at three depths $z\in\{\text{-}1.0,1.0,3.0\}\,\mu$m.
\proposed~demonstrates better sectioning capability than Tikhonov regularization. 
Superior removal of the missing-cone artifacts is also shown in the lateral views in Figure~\ref{Fig:diatom}(c) and~\ref{Fig:diatom}(d).
Because \proposed~learns a continuous representation of the RI distribution, it can generate images on arbitrarily dense voxel grids without additional retraining. Figure~\ref{Fig:diatom}(f) demonstrates this unique ability of \proposed~by interpolating $26.7\times$ more pixels in the $x$-$y$ planar region shown in Figure~\ref{Fig:diatom}(b).
For comparison, we used nearest neighbor (\emph{Pixel}), bilinear (\emph{Bilinear}), and bicubic (\emph{Bicubic}) interpolation methods to upsample the same region.
Our results show that \proposed~is able to resolve small features with strong cross-scale consistency while avoiding interpolation artifacts highlighted by the arrows.

\begin{figure}[t!]
\begin{center}
\includegraphics[width=\linewidth]{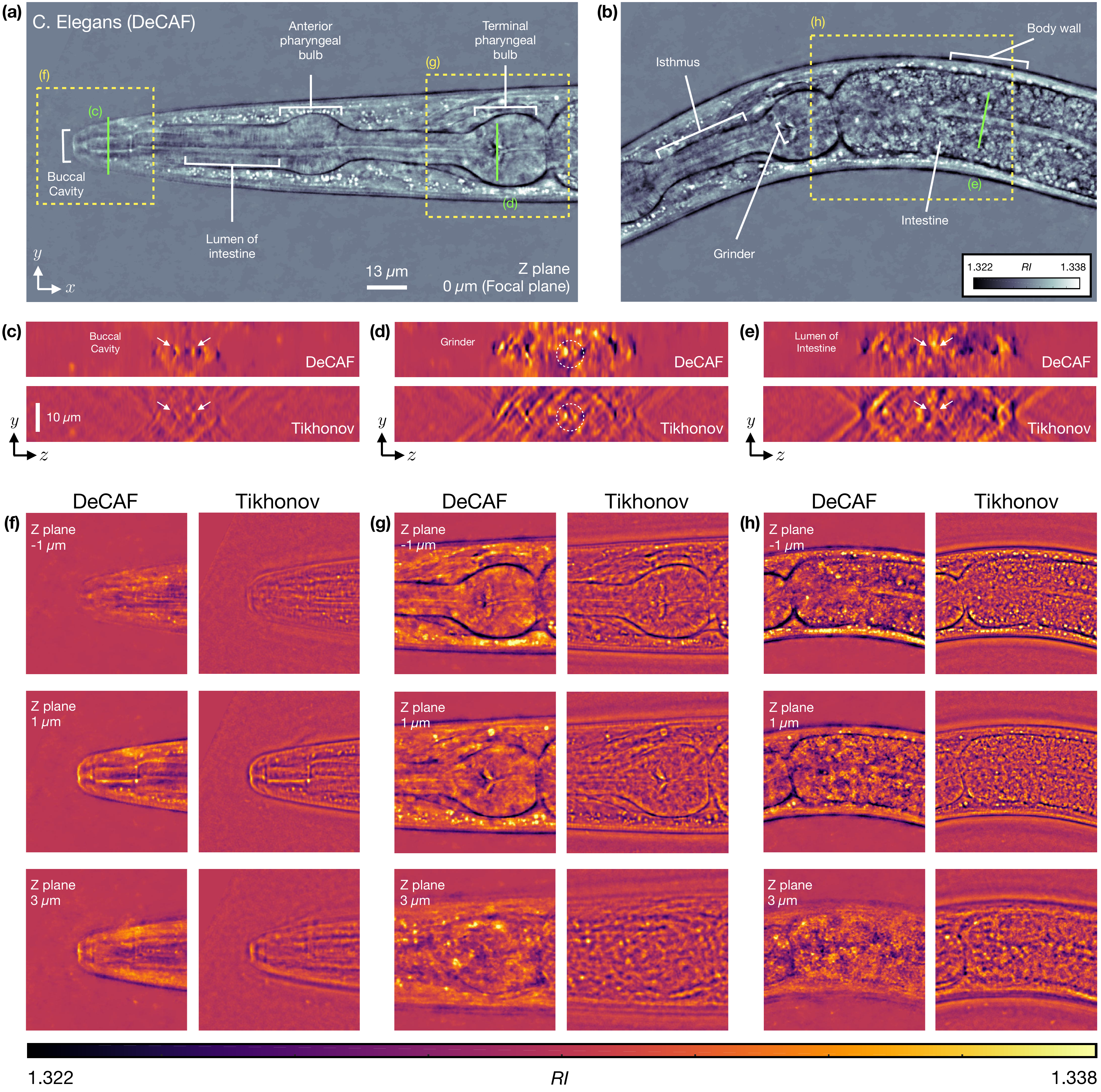}
\end{center}
\caption{
\textbf{Reconstruction of \emph{C.\ Elegans} acquired by mIDT.}
\textbf{(a) \& (b)} The reconstructed RI distribution at $z=0\,\mu$m (focal plane) by \proposed.
\textbf{(c), (d), \& (e)} Lateral views corresponding to the paths shown in (a) and (b). Biological structures are highlighted by the arrows and circles.
\textbf{(f), (g), \& (h)} Axial views of the regions highlighted in (a) and (b) at $z\in\{-1,1,3\}\,\mu$m. Note how \proposed~provides higher contrast and finer details than Tikhonov.
Additional examples are shown in Supplementary Videos \emph{celegans-decaf-head.mov}, \emph{celegans-decaf-body.mov}, \emph{celegans-tikhonov-head.mov} and \emph{celegans-tikhonov-body.mov}.
}
\label{Fig:celegans}
\end{figure}

We next present the results of epithelial buccal cell clusters in Figure~\ref{Fig:cells}. 
A background-removed (b.r.) intensity measurement showing the distribution of the whole cell cluster is presented in Figure~\ref{Fig:cells}(a).
We focused on two difficult regions where cells overlap with each other to highlight the superior axial sectioning capability of \proposed in Figure~\ref{Fig:cells}(b) and~\ref{Fig:cells}(c).
The size of the two volumes are $81.25\times81.25\times16\,\mu$m$^3$ and $97.5\times97.5\times16\,\mu$m$^3$, discretized to $32$ slices of $500 \times 500$ pixels and $600\times600$, respectively.
\proposed~successfully resolves different cells with clear separation while Tikhonov generates strong artifacts that blur the boundaries. Visual demonstration of the axial slices of these cells are provided in Figure~\ref{Fig:cells}(d)-\ref{Fig:cells}(f).
In each reconstructed slice, \proposed~recovers clear cell membrane, cytoplasm, micronuclei, and bacterias while removing the diffraction and scattering artifacts.

We further show the continuous representation learned by \proposed~by upsampling it along $z$, meaning that \proposed~is used to interpolate an entire axial slice that was not part of the grid used during training. Figure~\ref{Fig:cells}(h) and~\ref{Fig:cells}(i) present the interpolated slices of the bacteria clusters highlighted in Figure~\ref{Fig:cells}(d) and~\ref{Fig:cells}(e), respectively.
In each figure, a $z$ axis is provided to show the axial location of each slice. 
Note that $\{-5.5, -4.5, -3.5\}\,\mu$m and $\{1.5, 2.5, 3.5\}\,\mu$m are the axial coordinates pre-defined in the training grid.
The interpolated slices in Figure~\ref{Fig:cells}(h) and~\ref{Fig:cells}(i) clearly show the appearance and disappearance of the bacteria clusters at different values of $z$. 
As shown, the interpolated biological features are consistent with the ones lying in the pre-defined grid, making the whole transition smooth across axial layers.
This strong axial consistency preserved in \proposed~enable it to produce high-fidelity interpolations without any additional re-training. 

We finally validate \proposed~on the \emph{multiplexed IDT} microscopy (mIDT). This modality allows more rapid acquisition under the same time by simultaneously illuminating the sample from multiple angles for each intensity measurement. 
We imaged a Caenorhabditis elegans (C.\ elegans) worm specimen by using a $0.65$ NA objective lens to acquire $16$ measurements, with each from the simultaneous illuminations of $6$ different LED sources.
Figure~\ref{Fig:scheme}(b) (see images e-f) shows two example measurements.
The sample is challenging due to its thickness and complicated arrangement of organs. 
As the worm is live and moving during the acquisition, we reconstructed two volumes of $162.5\times162.5\times20\,\mu$m$^3$, discretized to $40$ slices of $1000\times1000$ pixels, at different times to cover the worm body with interested biological features.
Extended Data Figure~1 additionally shows the ability of \proposed~to reconstruct a relatively thin Diatom Algae sample from mIDT measurements.

Figure~\ref{Fig:celegans}(a) and~\ref{Fig:celegans}(b) present the reconstructed RI maps of C.\ elegans at the focal plane. 
\proposed~successfully recovers the sample's structure with clear quantification of the internal biological tissues. For example, the buccal cavity, anterior and terminal pharyngeal bulbs, isthmus, and intestine are clearly restored in our reconstruction. 
Smaller features are also distinguishable with high contrast, shown in the regions expanded in Figure~\ref{Fig:celegans}(f)-\ref{Fig:celegans}(h). 
For example, lysosomes, grinder, and lumen of intestine are accurately visualized with clear separation from the other tissues.
Figure~\ref{Fig:celegans}(c)-\ref{Fig:celegans}(e) show the $y$-$z$ lateral views, where the oval shape of C.\ elegans is reconstructed without the missing-cone artifacts, and fine features such as buccal cavity and grinder are preserved and recovered at different axial layers.

\begin{figure}[t!]
\begin{center}
\includegraphics[width=\linewidth]{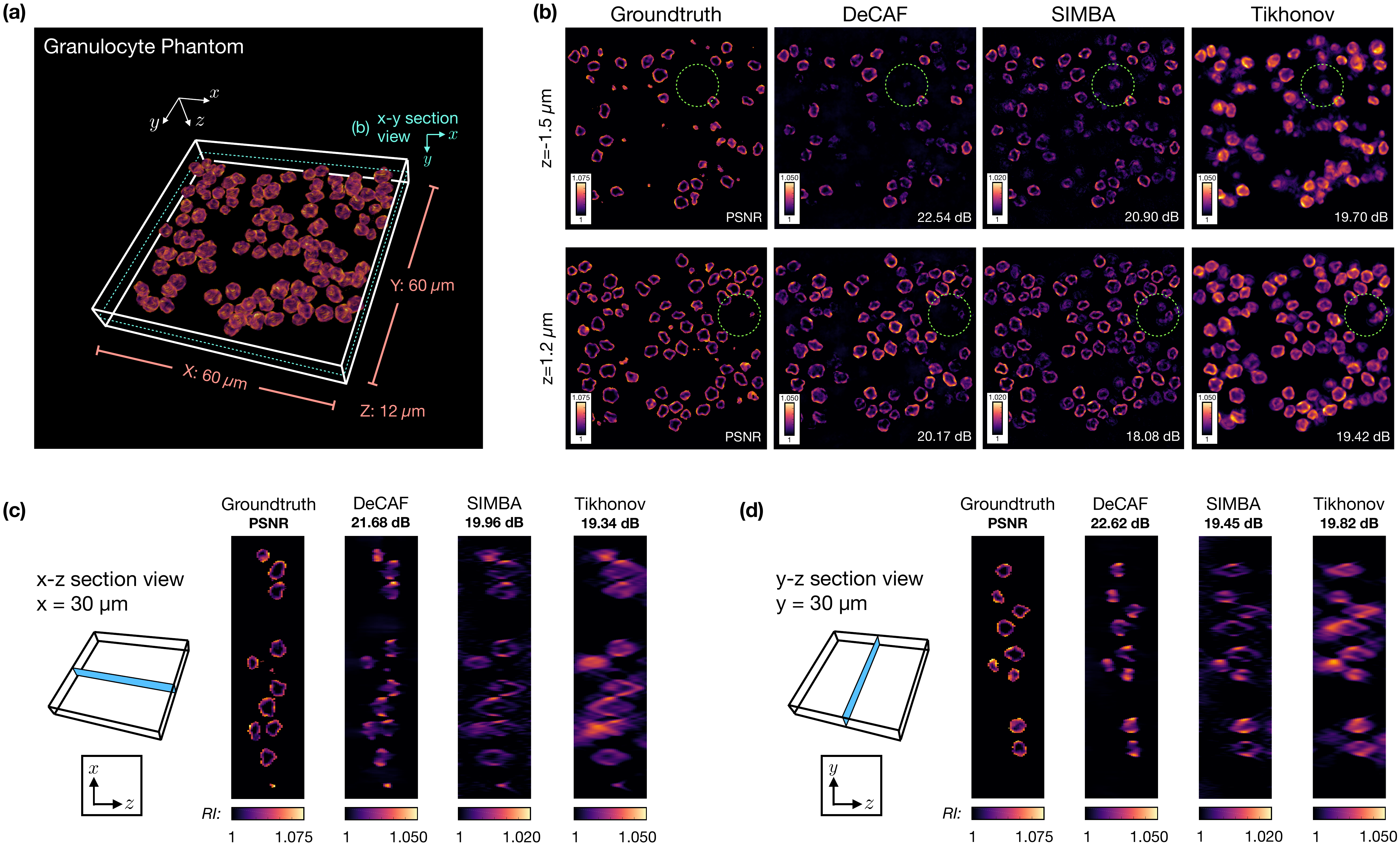}
\end{center}
\caption{
\textbf{Quantitative and visual comparison of \proposed, SIMBA, and Tikhonov for the reconstruction of the \emph{Granulocyte Phantom} from simulated data.} 
\textbf{(a)} 3D illustration of the phantom. 
\textbf{(b)} Visual comparison of the axial views at $z\in\{1.2, -1.5\}\;\mu$m reconstructed using all three methods. Groundtruth is provided in the leftmost column.
\textbf{(c) \& (d)} Visual comparison of the $x$-$z$ and $y$-$z$ lateral views reconstructed using each method. The corresponding position of each view is shown in the left of each figure.
Note how \proposed~provides much higher PSNR values than both SIMBA and Tikhonov.
}
\label{Fig:Granulocyte}
\end{figure}

Extended Data Figure~2 highlights the space used for storing the MLP weights in \proposed.
Since the representation is decoupled from a predefined voxel grid, \proposed~can be trained on a sparse grid to reduce the storage cost, but can still produce the final reconstruction on a grid of desired density. The storage reduction is demonstrated by comparing the memory requirements of \proposed~and Tikhonov for the reconstruction of the C. elegans worm. \proposed~retains the small memory size of $3$ MB across different grid densities while that of Tikhonov increases as the grid becomes denser.

\subsubsection*{Quantitative Evaluation}
In this section, we present quantitative evaluations of \proposed~using a high-fidelity cell phantom. 
We used CytoPacq~\cite{Wiesner.etal2019} to generate a \emph{granulocyte phantom} containing tens of granulocyte cells randomly distributed in a volume of $60\times60\times12\,\mu\text{m}^3$, discretized to $40$ slices of $454\times454$ pixels.
The maximum RI value in the cell is set to $\nbm_\text{re} = 1.075$ and $\nbm_\text{im} = 0$, meaning that there is no absorption.  The immersion media is assumed to be air ($\nbm_0=1$).
The simulation is based on the aIDT setup and used the \emph{split-step non-paraxial (SSNP)} simulator to simulate the full wave propagation~\cite{Lim.etal2019,zhu_intensity_2021}. The setup includes an annular LED array at $515$ nm wavelength for illumination and an objective lens with $0.65$ NA. During the acquisition,  in-total 24 measurements are taken. 

The result of quantitative evaluation are summarized in Figure~\ref{Fig:Granulocyte}. Figure~\ref{Fig:Granulocyte}(a) shows the overall 3D structure of the phantom.
Section views of axial and lateral planes are compared for \proposed, SIMBA, and Tikhonov, with quantitative evaluation of the \emph{peak signal-to-noise ratio (PSNR)} values 
\begin{equation}
\label{eq:PSNR}
\text{PSNR}(\xbm, \xbmhat) =  10\log\left(\frac{\max(\xbm)}{\mse(\xbm,\xbmhat)}\right),
\end{equation}
where $\mse(\cdot,\cdot)$ computes the \emph{mean squared error (MSE)}, and $\max(\cdot)$ returns the maximum pixel value in the image. Figure~\ref{Fig:Granulocyte}(b) visualizes the axial slices reconstructed by each method at the depths $z\in\{-1.5, 1.2\}\,\mu\text{m}$. 
Figure~\ref{Fig:Granulocyte}(c) and~\ref{Fig:Granulocyte}(d)  plot the $x$-$z$ and $y$-$z$ section views, respectively. The corresponding position of each view is shown in the left of each figure. \proposed~achieves better PSNR than both SIMBA and Tikhonov by reconstructing more accurate RI values, reducing the missing-cone artifacts, and removing the cell shadows due to axial elongations (highlighted using dashed circles). 

Extended Data Figure~3 visualizes the 3D volumes reconstructed by each method using Fiji~\cite{Schindelin:2012ty} under the default configuration. 
From left to right, the figure displays the 3D volumes corresponding to Groundtruth, \proposed, SIMBA, and Tikhonov. PSNR values are labeled on each volume in green.
\proposed~clearly outperforms SIMBA and Tikhonov by reconstructing cells that look most similar to Groundtruth. 
For example, consider the cells highlighted in the zoom-in volume. 
\proposed~reconstructs these cells with clear shapes and sharp edges, while the reconstructions of SIMBA and Tikhonov are either axially elongated or blurry.  Quantitative results further highlight the accuracy of \proposed, showing PSNR improvements of $1.6$ dB and $3.3$ dB with respect to SIMBA and Tikhonov, respectively (equivalent to $1.5\times$ and $2.1\times$ reduction in MSE). 

\subsection*{Discussion}
\textbf{Difference to SIMBA.} \proposed~offers several benefits from the existing SIMBA method: 
\emph{(i) Test-time learning (TTL):} SIMBA does \emph{not} adapt to the specifics of a test sample---it uses a fixed forward model and a fixed pre-trained prior. On the other hand, \proposed~is a TTL method where the MLP weights are adjusted for \emph{each} test sample, leading to a better reconstruction performance reported throughout this paper. 
\emph{(ii) Grid-free representation: } SIMBA reconstructs a discrete volume on a pre-defined voxel grid. \proposed~decouples the representation of the reconstructed 3D RI from the grid by using MLP. This enables one to synthesize any part of the 3D RI ``on demand'' on any grid by simply querying the relevant coordinates of MLP. Thus, the complexity of storing the sample reconstructed by \proposed~is decoupled from the voxel-grid.
\emph{(iii) Internal and external regularization: } Unlike SIMBA, \proposed~synergistically uses internal and external regularization offered by MLP and a CNN denoiser, respectively. Our quantitative results show that MLP offers significant amount of regularization, even when no external regularizer is used. The best results, however, are achieved when both regularizers are used.

\vspace{0.5em}
\noindent
\textbf{Limitations of \proposed.}
An obvious limitation of \proposed~is that it is based the linear IDT forward models based on the first Born approximation. This limits the applicability of the current implementation to relatively thin and weakly scattering samples. This limitation can be observed in the reconstruction of a relatively thick C.\ elegans sample. 
Future work will explore the extension of \proposed~to thicker and stronger scattering samples by using forward models accounting for multiple scattering, such as the ones based on the variations of the beam propagation method~\cite{Kamilov.etal2015, Tian.Waller2015}. 
Another limitation of \proposed~is that it is currently slower than existing IDT reconstruction methods, Tikhonov and SIMBA, which is due to our implementation of the NF training.
Approximately, our model takes less than a day ($\sim$ 20 hours) to infer each real sample, while the runtimes of Tikhonov and SIMBA are at the levels of several minutes and hours, respectively.
In addition, the hyperparameters of \proposed~need to be tuned manually on real samples due to the lack of groundtruth, which potentially leads to further increases in runtime.
The future work will explore faster \proposed~implementations that leverage recent progress in accelerating NF methods (for example, \emph{Instant Neural Graphics Primitives}~\cite{Mueller.etal2022} suggests an order of magnitude acceleration).

\subsection*{Conclusion}
\label{Sec:Conclusion}
We proposed a novel self-supervised deep learning method, \proposed, for enabling high-quality 3D reconstruction of the RI distribution from intensity-only measurements. 
We extensively validated \proposed~on the experimentally collected datasets of multiple biological samples under three different IDT setups. 
The results show that \proposed~can mitigate the missing-cone artifacts while maintaining the fine details of small biological features.
We additionally provide quantitative evidence to further corroborate our argument. Results show that \proposed~can reduce MSE by up to $2.1\times$.
The continuous representation in \proposed~also allows to generate images at voxel grids of arbitrary density without retraining of the deep network, which is useful for addressing computational and memory bottlenecks in image reconstruction and analysis.

\section*{Methods}
\label{Sec:Methods}
\subsection*{IDT Experiments}

\textbf{IDT Resolution:} The lateral and axial resolution of the IDT system is limited by the support of the optical transfer function (TF), which is determined by the objective NA and illumination NA~\cite{Ling.etal18}. For both aIDT and mIDT setup, our maximum illumination angle is close to the objective NA. Thus, the recovered lateral spatial frequency can reach the ``incoherent'' diffraction limit $4\text{NA}/\lambda$, and the axial Fourier coverage is up to $(2\nbm_0 - 2\sqrt{\nbm_0^2-\text{NA}^2})/\lambda$, where $\nbm_0$ is the RI of background media.

\vspace{0.5em}
\noindent\textbf{Dense IDT.}
Our dense IDT system consists of a Nikon TE 2000-U microscope equipped with a custom programmable LED array, approximately illuminating plane wave with central wavelength $\lambda=632$~nm; a $10\times$/$0.25$~NA objective (Nikon, CFI Plan Achromat), and an sCMOS camera (PCO.Edge 5.5).
The LED array is placed about $79$~mm away from the sample.
It is controlled via a microcontroller and is synchronized with the camera.
A small subset of the LEDs on the array, which contains the $89$ LEDs within the brightfield region, is used to illuminate the sample sequentially.

\vspace{0.5em}
\noindent\textbf{Annular IDT.} Our annular IDT system consists of a Nikon ECLIPSE E200 microscope equipped with a programmable ring LED unit (Adafruit, 1586 NeoPixel Ring). The microscope objective is $40\times$/$0.65$ NA (Nikon, CFI Plan Achromat), and each LED approximately provides a plane wave with central wavelength $\lambda=515$~nm.
The ring LED unit has 24 LED lights and is $60$~mm in diameter. It is centered at the optical axis and placed approximately $35$~mm away from the sample, which sets the angle between the wave vector and the optical axis to about $40^\circ$ and complies with the microscope objective NA\@. 

\vspace{0.5em}
\noindent\textbf{Multiplexed IDT.} Our multiplexed IDT system has the same hardware specification as the dense IDT system except that the microscope objective is $40\times$/$0.65$~NA (Nikon, CFI Plan Achromat).
Besides, the subset of the LEDs used in the experiment changes to $96$ LEDs corresponding to the NA range from $0.3$ to $0.575$.
This design contains $16$ disjoint illumination patterns and the multiplexed illumination quantity of each pattern is $6$.
The camera is synchronized with the LED array and captures $16$ measurements corresponding to each illumination pattern.

\subsection*{Sample and data preparation}
\noindent
\textbf{Spirogyra Algae.} This sample is a part of Fisher Science Education algae basic slide set S68786. We captured $89$ intensity-only bright field measurements. 
We pre-processed each measurement by removing the background intensity followed by normalization. 
The same pre-processing procedure is also applied to other samples.
We consider a reconstruction volume of $665.6 \times 665.6 \times 80\,\mu \text{m}^3$, positioned between $-30\,\mu$m and $50\,\mu$m around the focal plane. The volume is discretized into 40 slices along the $z$ axis, with each slice having $1024 \times 1024$ pixels. Here, a single voxel corresponds to $6.5 \times 6.5 \times 2 \,\mu$m$^3$. 

\vspace{0.5em}
\noindent\textbf{Diatom Algae (aIDT).}
This sample is a part of Fisher Science Education algae basic slide set S68786.
We captured $24$ measurements and consider a reconstruction volume of $113.75 \times 113.75 \times 26\,\mu$m$^3$, positioned between $-10\,\mu$m and $16\,\mu$m around the focal plane. The volume is discretized into 52 slices along the $z$ axis, with each slice having $700 \times 700$ pixels. Here, a single voxel corresponds to $0.1625 \times 0.1625 \times 0.5\,\mu$m$^3$.

\vspace{0.5em}
\noindent\textbf{Diatom Algae (mIDT).}
This sample is a part of Fisher Science Education algae basic slide set S68786.
We captured $16$ measurements, and each measurement used $6$ LED lights.  
We consider a reconstruction volume of $130 \times 130 \times 30\,\mu$m$^3$, positioned between $-15\,\mu$m and $15\,\mu$m around the focal plane. The volume is discretized into 60 slices along the $z$ axis, with each slice having $800 \times 800$ pixels. Here, a single voxel corresponds to $0.1625 \times 0.1625 \times 0.5\,\mu$m$^3$.

\vspace{0.5em}
\noindent\textbf{Human Buccal Epithelial Cells.} This sample was swabbed from a researcher's buccal.
The individual rinsed the mouth with clean water and then twirled a wooden swab against the inner cheek.
The end of the swab was immersed in a drop of purified water on a glass slide and covered by a coverslip. 
We captured $24$ measurements of the cell cluster and consider two volumes in the region as shown in Figure~\ref{Fig:cells}(b) and~\ref{Fig:cells}(c). The former has $81.25\times81.25\times16\,\mu$m$^3$ and the latter has $97.5\times97.5\times16\,\mu$m$^3$. Both volumes are positioned between $-8\,\mu$m and $8\,\mu$m around the focal plane. They are discretized to $32$ slices of $500 \times 500$ pixels and $600\times600$ pixels. Here, A single voxel corresponds to $0.1625 \times 0.1625 \times 0.5\,\mu$m$^3$.

\vspace{0.5em}
\noindent
\textbf{Caenorhabditis Elegans.} Young adult C. elegans were mounted on $3\%$ agarose pads in a drop of nematode growth medium (NGM) buffer. Glass coverslips were then gently placed on top of the pads and sealed with a 1:1 mixture of paraffin and petroleum jelly. 
As the C. elegans was alive and moving during data acquisition, we captured a video at 4 frames per second, where each frame contained 16 measurements, and each measurement used $6$ LED lights. 
We picked two frames at $1.5s$ and $44s$ for reconstruction, where the sample was relatively steady. We consider a unified reconstruction volume of $162.5 \times 162.5 \times 20\,\mu \text{m}^3$, positioned between $-10\,\mu$m and $10\,\mu$m around the focal plane. 
The volume is divided into $40$ slices along the $z$ axis, with each slice having $1000 \times 1000$ pixels. Here, a single voxel corresponds to $0.1625 \times 0.1625 \times 0.5\,\mu$m$^3$.

\subsection*{\proposed~Framework}
A linearized approximation of IDT forward measurement system can be described by equation~\eqref{Eq:linearIDT}
\begin{equation}
\label{Eq:linearIDT}
\ybm_\rho \approx \Abm_\rho \Delta \bm{\epsilon},
\end{equation}
where $\Delta \epsilonbm = \Delta \epsilonbm_\text{re} + j\Delta \epsilonbm_\text{im}$ is the unknown volume of complex-valued permittivity contrast, $\ybm_\rho$ is the collection of the background-removed intensity measurements corresponding to the LED illuminations emitted at a set of locations $\rho$, and $\Abm_\rho$ is the measurement matrices that model the sample-intensity mapping associated with these illuminations.
The reconstruction of $\Delta \epsilonbm$ is equivalent to the reconstruction of the RI distribution via equation~\eqref{Eq:translation}
\begin{equation} \label{Eq:translation}
\bm{n}_\text{re} = \sqrt{\frac{1}{2}((\bm{n_0}^2 + \Delta \epsilonbm_\text{re}) + \sqrt{(\bm{n_0}^2+\Delta \epsilonbm_\text{re})^2 + \Delta \epsilonbm_\text{im}^2})} \quad\text{and}\quad \bm{n}_\text{im} = \frac{\Delta \epsilonbm_\text{im}}{2\cdot \bm{n}_\text{re}},
\end{equation}
where $\bm{n}_\text{re}$ and $\bm{n}_\text{im}$ are the real and imaginary parts of the sample's RI, and $\bm{n}_0$ is the RI of the background medium (where the attenuation is often assumed to be zero). In equation~\eqref{Eq:translation}, all operations are evaluated in an element-wise manner. 
We derived the formulations of $\Abm_\rho$ by following the prior work on dIDT~\cite{Ling.etal18}, aIDT~\cite{LiJ.etal2019}, and mIDT~\cite{Matlock.etal2019} (see \emph{IDT Forward Model} in the Supplement).

The central piece of \proposed~is a coordinate-based MLP, $\Mcal_\phi$, which maps the 3D coordinate $(x,y,z)$ to the corresponding values of $\Delta \epsilon_\text{re}$ and $\Delta \epsilon_\text{im}$.
We normalize the coordinate grid to a cube $[-1,1]^3$ before feeding them into $\Mcal_\phi$.
The deep network $\Mcal_\phi$ consists of two subnetworks, where the first one is an encoding layer $\gamma(x,y,z)$, pre-defined before training, and the second one is a standard MLP $\Ncal_\phi:\gamma(x,y,z)\rightarrow (\Delta \epsilon_\text{re}, \Delta \epsilon_\text{im})$ parameterized by the trainable parameters $\phi$.  A visual illustration of the detailed network architecture is provided in the Extended Data Figure~4(a).

\vspace{0.5em}
\noindent
\textbf{Radial Encoding.} It has been shown that a Fourier-type encoding of the spatial coordinates is essential for a MLP to represent high-frequency variations in the signal~\cite{Mildenhall.etal2020} and impose implicit regularization~\cite{Sun.etal2021a}.
In \proposed, we consider a decomposition of the input coordinate $(x,y,z)$ into $(x,y)$ and $z$, and use different strategies to expand $(x,y)$ and $z$.
This is due to the non-isotropic resolution of the imaging system along the $x$-$y$ plane and the $z$ dimension.  Our experiments showed that existing encoding strategies, such as positional~\cite{Mildenhall.etal2020} and Gaussian~\cite{Tancik.etal2020} encoding, lead to  suboptimal reconstruction of RI images along the $x$-$y$ dimensions.
We propose \emph{radial encoding} as an alternative for expanding $\vbm\defn(x,y)$
\begin{align}
\label{Eq:Rad}
\gamma_\text{rad}(\vbm) = 
        \begin{pmatrix}
           \sin\left(2^0\pi\Rbm_{\bm{\theta}}\vbm\right),\cos\left(2^0\pi\Rbm_{\bm{\theta}}\vbm\right), \\
           \vdots \\
           \sin(\underbrace{2^{L_{xy}-1}}_{k_\text{sin}}\pi\Rbm_{\bm{\theta}}\vbm),\cos(\underbrace{2^{L_{xy}-1}}_{k_\text{cos}}\pi\Rbm_{\bm{\theta}}\vbm)
         \end{pmatrix}\quad\text{with}\quad\Rbm_{\bm{\theta}}=
         \left\{\begin{bmatrix}
           \cos(\theta_k)& -\sin(\theta_k)\\
           \sin(\theta_k) & \cos(\theta_k)
         \end{bmatrix}\right\}_{k=1}^K.
\end{align}
Here, $\sin$ and $\cos$ compute the (element-wise) sinusoidal and cosinusoidal values, respectively, $\Rbm_{\bm{\theta}}$ denotes a collection of rotation matrices that translate the coordinate by the angles ${\bm{\theta}}$, and $L_{xy}>0$ controls the number of the expanded frequency.
By incorporating rotation, our strategy enables a frequency expansion that can efficiently acount for the dependencies within the $x$-$y$ plane (see \emph{Radial Encoding} in the Supplement).
The difference between radial encoding and positional encoding is conceptually illustrated in Extended Data Figure~4(b) and Extended Data Figure~4(c). In the experiments, we observed that the radial encoding improves the representation of small textures that are otherwise lost by other encodings. We adopted the standard positional encoding for the expansion of $z$
\begin{align}
\label{Eq:Pos}
\gamma_\text{pos}(z) = 
        \begin{pmatrix}
           \sin\left(2^0\pi z\right),\cos\left(2^0\pi z\right), \\
           \vdots \\
           \sin(\underbrace{2^{L_z-1}}_{k_\text{sin}}\pi z),\cos(\underbrace{2^{L_z-1}}_{k_\text{cos}}\pi z)
         \end{pmatrix},
\end{align}
where $L_z>0$ denotes the total number of frequencies.
We fine-tuned $\bm{\theta}$, $L_{xy}$, and $L_{z}$ for every sample by running multiple sets of parameters and manually selecting the set leading to the best visual quality. 
We summarize their values in Extended Data Figure~5.
The ablation experiment on the challenging C.\ elegans specimen (see details in Supplementary Figure 4) demonstrates the superior performance of the proposed encoding. Additional quantitative evidence is also presented in \emph{Ablation Experiments} in the Supplement.

\vspace{0.5em}
\noindent
\textbf{MLP architecture.}
The network architecture of $\Ncal_\phi$ is illustrated in Extended Data Figure~4(a).
Network $\Ncal_\phi$ is composed of $N$ fully-connected (FC) layers.
The first $N-1$ layers have $M$ hidden neurons activated by the leaky rectified linear unit (Leaky ReLU), while the last layer has $M$ unactivated hidden neurons.
A skip connection is implemented at the $\lfloor N/2 \rfloor^\text{th}$ FC layer to concatenate the original input of $\Ncal_\phi$ with the intermediate outputs, which has been shown beneficial for improving the representation performance~\cite{Park.etal2019}.
We used one network configuration for all biological samples, which is summarized in Extended Data Figure~5.

\vspace{0.5em}
\noindent
\textbf{Regularized loss function.}
At test time, we train $\Mcal_\phi$ to minimize equation~\eqref{Eq:loss} by using a customized Adam~\cite{Kingma.Ba2015} optimizer (see \emph{Block-wise Training of \proposed} in the Supplement)
\begin{equation}
\label{Eq:loss}
\Lcal(\phi;\ybm_\rho) = \underbrace{\|\Abm_\rho(\Mcal_\phi(\cbm)) - \ybm_\rho\|_1}_\text{measurement consistency} + \alpha\underbrace{\|\Mcal_\phi(\cbm)-\Dsf_\sigma(\Mcal_\phi(\cbm))\|_2^2}_\text{$x$-$y$ plane noise reduction} + \beta\underbrace{\sum_j \|\Mcal^{j}_\phi(\cbm) - \Mcal^{j-1}_\phi(\cbm)\|_1}_\text{$z$-dimension continuity},
\end{equation}
where $\cbm=\{(x_i,y_i,z_i)\}_{i=1}^n$ is a collection of all coordinates on the grid, and $\Mcal_\phi^j$ denotes the $j$th axial slice of the predicted RI map.
The loss defined in equation~\eqref{Eq:loss} can be divided into three terms serving different purposes, with $\alpha\geq0$ and $\beta\geq0$ balancing their contributions.
The first term is a widely-used $\ell_1$-norm loss that ensures the consistency with the test measurements. 
The second and third terms are the regularizers imposing $x$-$y$ plane noise reduction and continuity along $z$, respectively.
$\Dsf_\sigma$ denotes a $2$D image denoiser with $\sigma>0$ controlling the denoising strength.
We selected DnCNN as our denoiser due to its state-of-the-art denoising performance~\cite{Zhang.etal2017}. 
A detailed description of the architecture and training of DnCNN is presented in \emph{Additional Technical Details} in the Supplement.
We fine-tuned $\alpha$ and $\beta$ for each sample by running multiple sets of parameters and manually selecting the set leading to the best visual quality. 
We summarized their values in Extended Data Figure~5
In the Supplement, we provide the empirical evidence to demonstrate the necessity of the explicit regularization for imaging the complex organism (see visual results in Supplementary Figure~6). 
Additional quantitative evaluations are provided in \emph{Ablation Experiments} in the Supplement.

\subsection*{Data availability}
The data used for reproducing the results in the manuscript is available at \url{https://github.com/wustl-cig/DeCAF} (ref.~\cite{Code_Data}). 
We visualized the pre-processed raw intensity images of the relevant samples in Figure 1 and Figure 4.

\subsection*{Code availability}
The code used for reproducing the results in the manuscript is available at \url{https://github.com/wustl-cig/DeCAF} (ref.~\cite{Code_Data}). 

\section*{Acknowledgement}
This work was supported by the NSF awards CCF-1813910 (U.K.), CCF-2043134 (U.K.), CCF-1813848 (L.T.), and EPMD-1846784 (L.T.).

\section*{Contributions}
The project was conceived by Y.S., R.L., and U.K.. The code of the model was implemented by R.L. and Y.S.. The experiments were designed by R.L. and Y.S.. The numerical results were collected by R.L.. The data acquisition and preparation was conducted by J.Z. and L.T.. The manuscript was primarily drafted by Y.S., assisted by R.L. and J.Z.. The manuscript was revised by U.K. and L.T., and reviewed by all authors.

\section*{Competing interests}
The authors declare no competing interests.

\newpage
\setcounter{figure}{0}
\renewcommand{\figurename}{Extended Data Figure}
\renewcommand{\tablename}{Extended Data Table}

\begin{figure}[t!]
\begin{center}
\includegraphics[width=\linewidth]{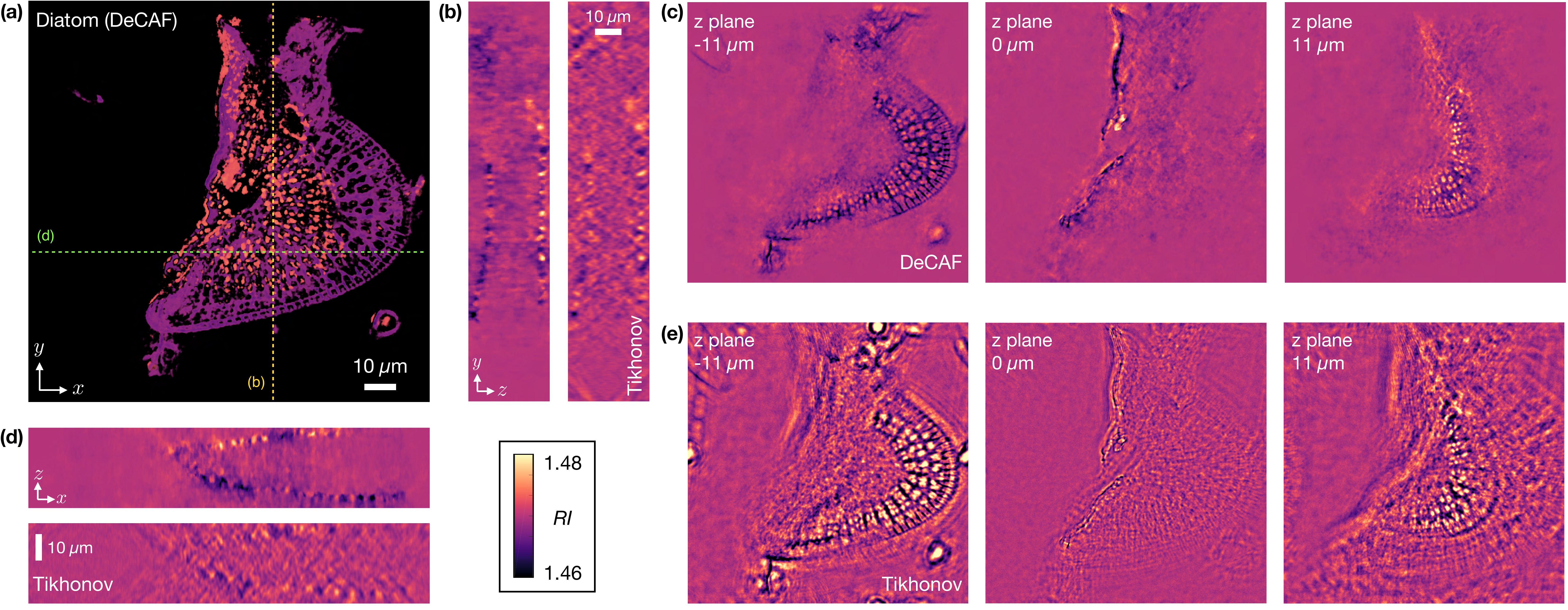}
\end{center}
\caption{
\textbf{Reconstruction of \emph{Diatom Algae} acquired by mIDT.} \textbf{(a)} 2D rendering obtained by accumulating all the $z$ slices from \proposed. Scale bar 10 $\mu$m.
\textbf{(b) \& (d)} Lateral views corresponding to the colored lines in (a). 
\textbf{(c) \& (e)} Axial views at $z\in\{-11, 0, 11\}\;\mu$m reconstructed by using \proposed~and Tikhonov, respectively.
This figure illustrates the ability of \proposed~to reconstruct high-contrast RI maps for a relatively thin sample acquired by mIDT.
Note how \proposed~successfully recovers the folding structure of the sample with two clear separate layers, which are barely recognizable in the Tikhonov reconstruction.  
Additional examples are shown in Supplementary Videos \emph{diatom-midt-decaf.mov} and \emph{diatom-midt-tikhonov.mov}.}
\label{Fig:DiatomMIDT}
\end{figure}
\clearpage

\newpage
\begin{figure}[t!]
\begin{center}
\includegraphics[width=\linewidth]{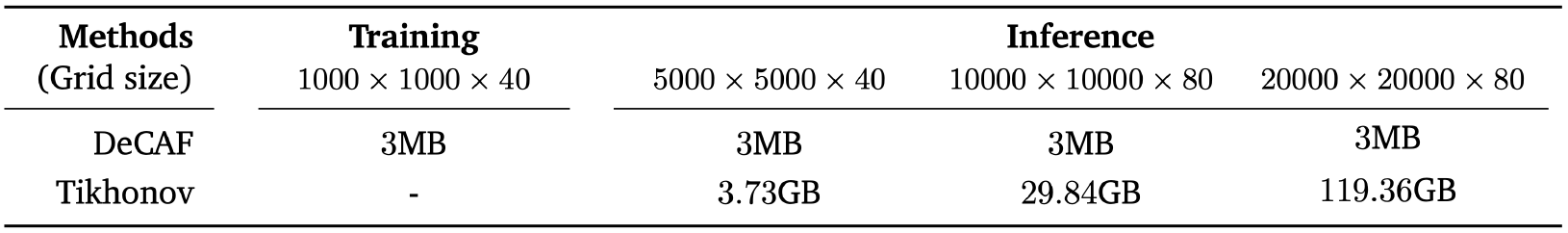}
\end{center}
\caption{Quantitative Illustration of the scalability of \proposed~due to its off-the-grid feature using the C.\ elegans specimen as an example. Note how the space required to store the reconstructed sample in \proposed~is independent of the reconstruction grid.}
\label{Tab:storage}
\end{figure}
\clearpage

\newpage
\begin{figure}[t!]
\begin{center}
\includegraphics[width=\linewidth]{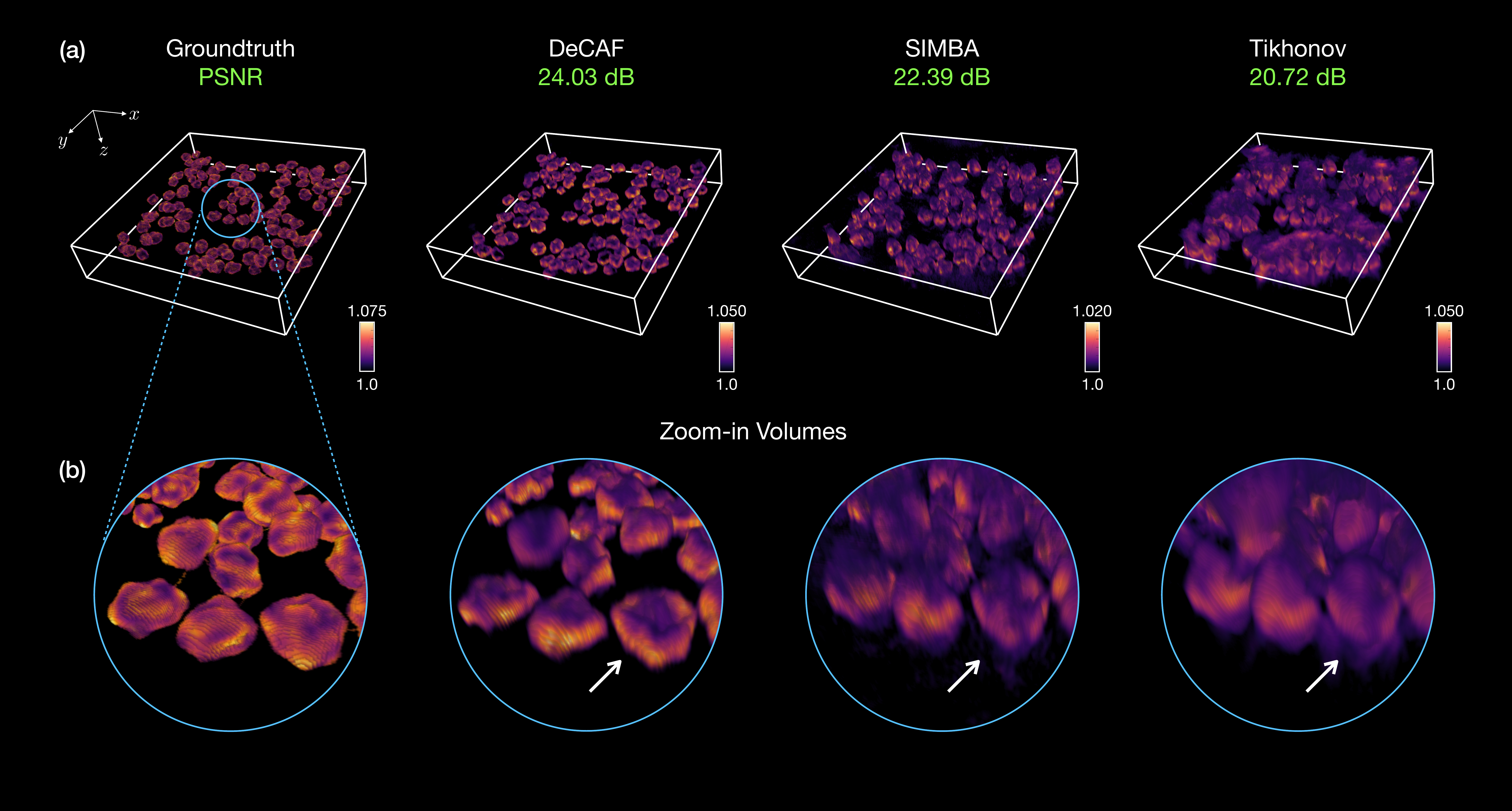}
\end{center}
\caption{\textbf{Reconstruction of the 3D \emph{Granulocyte Phantom} using \proposed, SIMBA, and Tikhonov.} \textbf{(a)} From left to right, 3D volumes correspond to Groundtruth, DeCAF, SIMBA, and Tikhonov, respectively. \textbf{(b)} Close-up views of the reconstructions at the location shown in (a). Note how DeCAF reconstructs sharper and better quality cell images compared to both SIMBA and Tikhonov.}
\label{Fig:Granulocyte3D}
\end{figure}
\clearpage

\newpage
\begin{figure}[t!]
\begin{center}
\includegraphics[width=\linewidth]{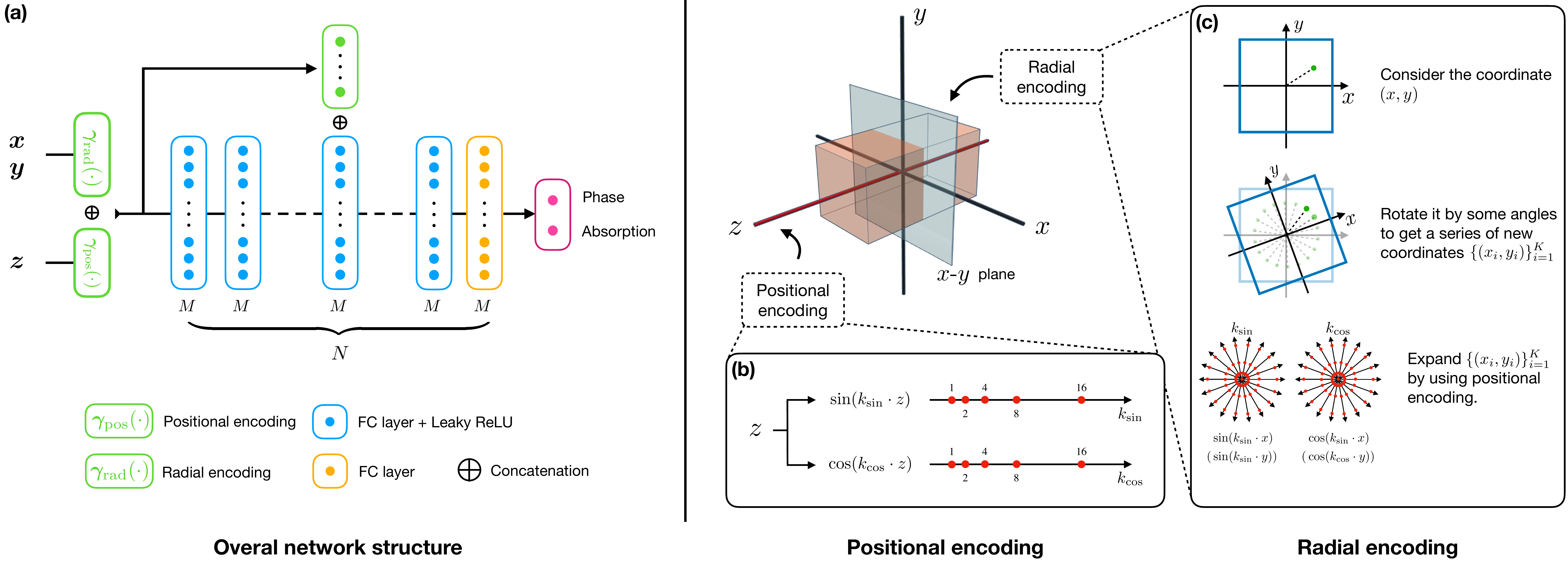}
\end{center}
\caption{
\textbf{Visual illustration of the network structure and the encoding strategy used in \proposed.}
\textbf{(a)} The overall structure of network $\Mcal_\phi$.
\textbf{(b)} Illustration of \emph{positional encoding} for $z$ coordinate.
\textbf{(c)} Illustration of \emph{radial encoding} for the coordinates in the $(x,z)$ plane.}
\label{Fig:network_radial}
\end{figure}
\clearpage

\newpage
\begin{figure}[t!]
\begin{center}
\includegraphics[width=\linewidth]{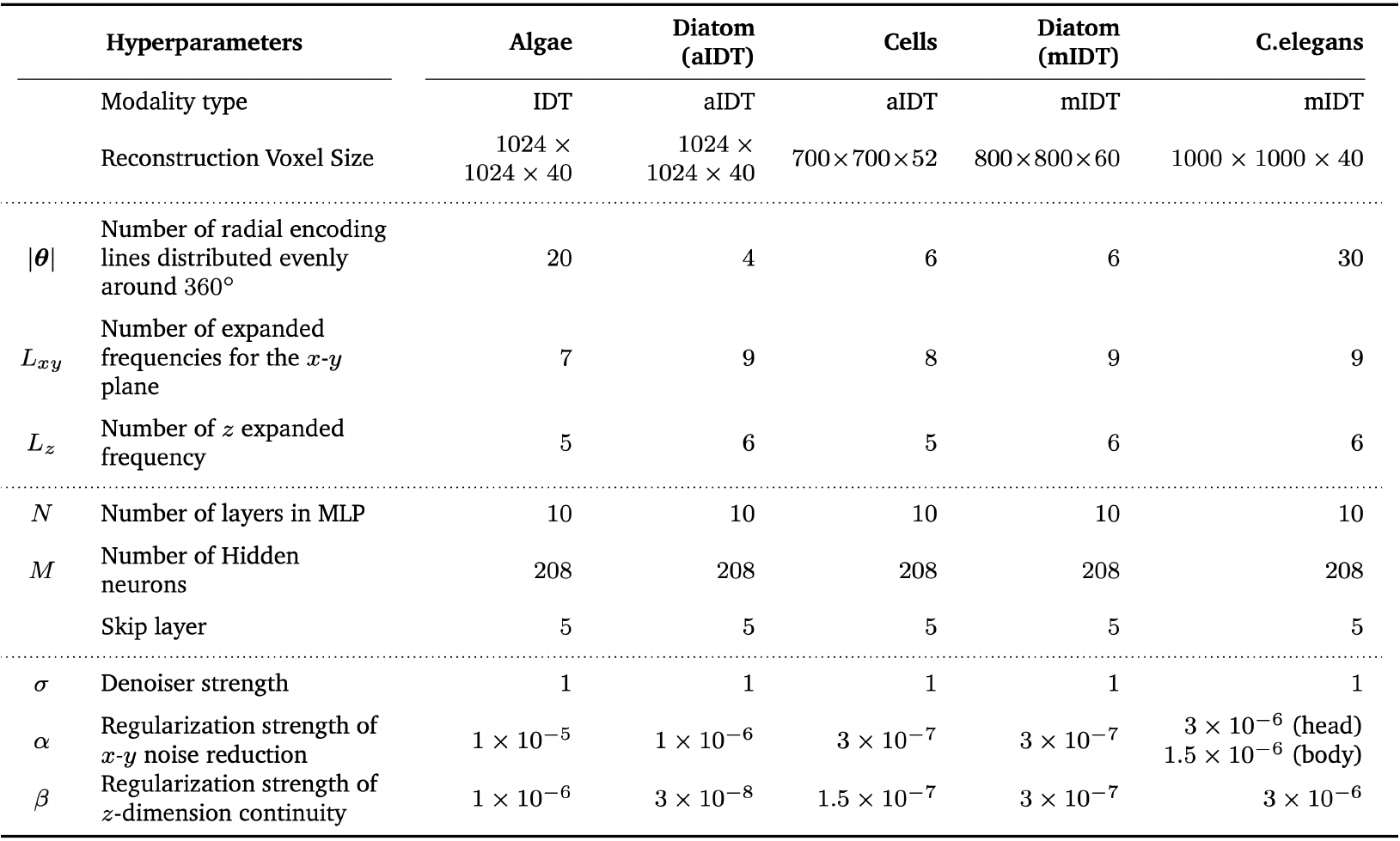}
\end{center}
\caption{
List of algorithmic hyperparameters used by \proposed~for different biological samples
}
\label{Tab:ListOfParams}
\end{figure}

\newpage

{\centering\LARGE \textbf{Supplementary Material for ``Recovery of Continuous 3D Refractive Index Maps from Discrete Intensity-Only Measurements using Neural Fields"}}

\section*{IDT forward model}

We adopt the traditional IDT forward model where the scattering events throughout the sample volume are characterized using the first Born approximation~\cite{Wolf1969}.
Hence, a 3D sample is discretized into a stack of 2D slices along the axial direction, and an individual measurement matrix is derived for each slice to linearly map the 2D permittivity contrast to the intensity measurements.
Given the $p^\text{th}$ illumination, the discrete IDT system is expressed as
\begin{equation}
\ybm_p = \sum_{q=1}^Q \Abm_{p,q}\Delta\epsilonbm_q + \ebm_0,
\end{equation}
where $\ybm_p\in\R^m$ is the intensity measurement with the background intensity removed, $\Delta\epsilonbm\in\C^{n}$ is permittivity contrast, $q=1,\dots,Q$ is the index of the axial slices, $\{\Abm_{p,q}\}$ are the measurement matrices, and $\ebm_0$ represent the noise.
From the IDT principle~\cite{Ling.etal18}, every $\Abm_{p,q}$ is a composition of $2$D (inverse) Fourier transform $\mathscr{F}$ ($\mathscr{F}^{-1}$) and a transfer function (TF) $\Hbm_{p,q}$ 
\begin{equation}
\Abm_{p,q} = \mathscr{F}^{-1} \Hbm_{p,q} \mathscr{F}.
\end{equation}
From the equation, we note that $\Hbm_{p,q}$ fully characterizes the permittivity-intensity mapping. 
Next, we describe how to compute the transfer function for each IDT modality used in our experiments.

\vspace{0.5em}
\noindent
\textbf{Dense and annular IDT.} In dense and annular IDT, the LED is turned on in a one-by-one manner. 
For a single-LED plane wave illumination, the analytical expression of the phase and absorption TFs~\cite{Ling.etal18,LiJ.etal2019} can be expressed as 
\begin{subequations}
\begin{align}\label{Eq:phaseTF}
H^\text{ph}_{p,q}(\ubm) =  \bm{j}\frac{k_0^2}{2}S(\ubm_p)
&\left( P^\ast(\ubm)P(\ubm-\ubm_p)\frac{\exp\{-\bm{j}[\eta(\ubm-\ubm_p)-\eta_i]\cdot q\Delta z\}}{\eta(\ubm-\ubm_p)} - \right. \nonumber\\
&\left.\;\; P(-\ubm_p)P^\ast(\ubm+\ubm_p)\frac{\exp\{\bm{j}[\eta(\ubm+\ubm_p)-\eta_i]\cdot q\Delta z\}}{\eta(\ubm+\ubm_p)}
\right)
\end{align}
\begin{align}\label{Eq:absorTF}
H^\text{ab}_{p,q}(\ubm) =
-\frac{k_0^2}{2}S(\ubm_p)
&\left( P^\ast(\ubm_p)P(\ubm-\ubm_p)\frac{\exp\{-\bm{j}[\eta(\ubm-\ubm_p)-\eta_i]\cdot q\Delta z\}}{\eta(\ubm-\ubm_p)} - \right. \nonumber\\
&\left.\;\; P(-\ubm_p)P^\ast(\ubm+\ubm_p)\frac{\exp\{\bm{j}[\eta(\ubm+\ubm_p)-\eta_i]\cdot q\Delta z\}}{\eta(\ubm+\ubm_p)}
\right)
\end{align}
\end{subequations}
where $\ubm$ denotes the lateral wave vector, $\bm{j}$ is the imaginary unit, $k_0 = 2\pi/\lambda$ is the wave number, $\lambda$ is the illumination wavelength, $S$ is the source function, $P$ is the objective pupil function, $P^\ast$ is the conjugate transpose of $P$, $\ubm_i$ is the $i^\text{th}$ lateral illumination wave vector, $\eta(\ubm) = \sqrt{k^2_0-|\ubm|^2}$ is the axial wave vector, $\eta_p(\ubm_p) = \sqrt{k^2_0-|\ubm_p|^2}$ is the illumination axial wave vector, and $\Delta z$ is the axial sampling spacing (i.e., slice spacing). 

We obtain the matrix expression of these TFs by performing discretization and normalization: $\Hbm^\text{ph}_{p,q} = H^\text{ph}_{p,q}/I_p$ and $\Hbm^\text{ab}_{p,q} = H^\text{ph}_{p,q}/I_p$, where $I_p$ denote the squared modulus of the incident light field. 
Here, the division denotes element-wise operation. 
A full expression of $\Hbm_p$ for the $p^\text{th}$ illumination is given as
\begin{equation}
\Hbm_{p} = 
\begin{pmatrix}
\Hbm^\text{ph}_{p,1}\quad \cdots \quad \Hbm^\text{ph}_{p,Q}\\ \\
\Hbm^\text{ab}_{p,1}\quad \cdots \quad \Hbm^\text{ab}_{p,Q}
\end{pmatrix}.
\end{equation}

\vspace{0.5em}
\noindent
\textbf{Multipixeled IDT.} We consider the intensity measurement of a multipixeled illumination $\ybm^\text{mul}_p$ as a summation of the measurements of individual sub-illuminations $\ybm^\text{sub}_p$~\cite{Matlock.etal2019}. The forward model in the frequency domain can be expressed as
\begin{equation}\label{Eq:mIDT}
\ybmhat^\text{mul}_p = \sum_{w_p=1}^W\ybmhat^\text{sub}_{w_p} = \underbrace{\sum_{w_p=1}^W\;\sum_{q=1}^Q\;\Hbm_{w_p,q}}_{\text{$\defn\Hbm^\text{mul}_p$ for mIDT}}\Delta\widehat{\epsilonbm}_q + \widehat{\ebm},
\end{equation}
where $\widehat{\cdot}$ denotes the vector in the Fourier space, $W>0$ is the total number of sub-illuminations, and $\Hbm^\text{mul}_p$ denotes the TF of a single multipixeled illumination. By regrouping and reordering the TFs of each sub-illumination, we can derive the formulation of $\Hbm^\text{mul}_p$ as follows
\begin{equation}
\Hbm^\text{mul}_p = 
\begin{pmatrix}
\Hbm^\text{ph}_{1,1}&\cdots&\Hbm^\text{ph}_{1,Q}\\
\vdots& \ddots &\vdots\\
\Hbm^\text{ph}_{W,1}&\cdots&\Hbm^\text{ph}_{W,Q}\\ \\
\Hbm^\text{ab}_{1,1}&\cdots&\Hbm^\text{ab}_{1,Q}\\
\vdots& \ddots &\vdots\\
\Hbm^\text{ab}_{W,1}&\cdots&\Hbm^\text{ab}_{W,Q}
\end{pmatrix}.
\end{equation}

\section*{Block-wise Training of \proposed}
Although the memory footprint of the MLP in~\proposed~is small, its optimization involves the computation of the IDT forward model, which scales linearly by the total of number of the input coordinates, and thus can easily excess the memory limit of a graphic processing unit (GPU). To address this, we proposed a block-wise training approach to reduce the GPU memory usage during optimization.

\vspace{0.5em}
\noindent
\textbf{General block-wise procedure.}
The block-wise procedure (see Supplementary Figure~\ref{Fig:training} for visual illustration) can be described as follows: During optimization, we horizontally divide the 3D grid into different blocks. 
At each time, \proposed~randomly selects a single block and predicts its RI volume. Then, the volume is used to compute the measurement mismatch (by using the IDT forward model) and evaluate the regularizer. The weights $\phi$ of the MLP are then updated by leveraging the existing deep learning optimization algorithm.

\vspace{0.5em}
\noindent
\textbf{Padding, view enlargement, \& measurement separation.} In order to correctly account for the IDT forward models that relies on the Fourier transform, our training procedure uses three computational techniques, that is, \emph{padding}, \emph{view enlargement}, and \emph{measurement separation}. 
\emph{Padding} is used to prevent boundary artifacts by letting each coordinate block overlap with others by a small region $p>0$ (see Supplementary Figure~\ref{Fig:training}(b)). \emph{View enlargement} horizontally enlarging the predicted RI block to consider a larger 2D region of the measurement (see Supplementary Figure~\ref{Fig:training}(c)), which ensures a sufficiently large field of view to cover the intensity information. 
\emph{Measurement separation} separates and uses the partial measurements associated only with the predicted RI block to impose measurement consistency, by cropping the corresponding measurement block and removing cross-block interference (see Supplementary Figure~\ref{Fig:training}(d)).
Note that our measurement separation technique can be efficiently implemented inside the training algorithm for \proposed.

\renewcommand{\algorithmiccomment}[1]{\bgroup\hfill{$\triangleright$~#1}\egroup}
\begin{algorithm}[t]
\caption{Block-wise Adam}
\label{Alg:blockwise}
\begin{algorithmic}[1]
\STATE \textbf{input: } Measurement $\ybm_\rho$, dictionary $\Psi$ tracking the partial measurements $\{\ybm_i\}_{i=1}^B$.
\STATE \textbf{Initialization:} $\Psi = \bm{0}$.
\FOR{$k=1,2,3,\dots$}
\STATE Randomly pick one block $i\in\{1,...,B\}$
\STATE $\ybm_i \leftarrow \Usf_i^\Tsf(\ybm_\rho-\sum_{j\neq i}\Usf_j\ybm_j)$ \COMMENT{Measurement separation}\\
\quad\quad where $\ybm_j \leftarrow \text{read}(\Psi, j)$ 
\STATE $\phi^{k+1} \leftarrow \texttt{AdamGradientUpdate}\{\Lcal(\phi^{k}; \ybm_i)\}$ \COMMENT{Backward propagation}
\STATE $\ybm^\text{new}_i \leftarrow \Abm_\rho \Mcal_{\phi^{k+1}}(\cbm_i)$ \COMMENT{Forward propagation} \\
\quad\quad where $\cbm_i$ denotes the coordinates of block $i$
\STATE $\Psi \leftarrow$ update($i, \ybm_i^\text{new}$)
\ENDFOR
\end{algorithmic}
\end{algorithm}

\vspace{0.5em}
\noindent
\textbf{Block-wise Adam algorithm.}
We customized Adam to realize the block-wise training procedure.
Consider the approximation of measurement $\ybm_\rho$ by the partial measurements $\{\ybm_i\}_{i=1}^B$ generated by $B>0$ padded and view-enlarged RI blocks
\begin{equation}
\label{Eq:approxY}
\ybm_\rho = \sum_{i=1}^B\Usf_i\ybm_i, \quad\text{where}\quad\ybm_i\in\R^{m_b},\quad\Usf_i:\R^{m_b}\rightarrow\R^{m},\quad\Usf_i^\Tsf:\R^{m}\rightarrow\R^{m_b}
\end{equation}
where we introduce the operator $\Usf_i: \R^{n_i} \rightarrow \R^n$ that injects a vector in $\R^{m_b}$ into $\R^n$ and its transpose $\Usf_i^\Tsf$ that extracts the $i$th block from a vector in $\R^n$. Note that the two operators will not change the values in the vector. 
Pre-defining blocks allows us to enable efficient measurement separation by reusing the intermediate results.
At every training step, the partial measurement of block $i$ can be computed by evaluating equation~\eqref{Eq:getParitalMeas}
\begin{equation}
\label{Eq:getParitalMeas}
\ybm_i = \ybm_\rho-\sum_{j\neq i} \ybm_j,
\end{equation}
where $\ybm_j$ denotes the partial measurement of a block other than $i$. By leveraging a dictionary $\Psi$ to track the latest partial measurements of each block $\{\ybm_i\}_{i=1}^B$, one can efficiently compute equation~\eqref{Eq:getParitalMeas} over training. 
Algorithm~\ref{Alg:blockwise} summarizes the algorithmic details of the customized block-wise Adam algorithm.
Note that the algorithm only requires the measurements of the testing sample as input, and the Adam optimizer can be replaced by other training optimizers.

\section*{Radial Encoding}
The proposed radial encoding can be viewed as a reconciliation of positional encoding based on the classic Fourier series expansion \begin{align}
\label{Eq:Pos}
\gamma_\text{pos}(\vbm) = 
        \begin{pmatrix}
           \sin\left(2^0\pi \vbm\right),\cos\left(2^0\pi \vbm\right), \\
           \vdots \\
           \sin(2^{L-1}\pi \vbm),\cos(2^{L-1}\pi \vbm)
         \end{pmatrix}, \quad\text{with}\quad\vbm=(x,y,z)
\end{align}
and Gaussian encoding that relies on a completely random mapping
\begin{align}
\label{Eq:Pos}
\gamma_\text{gau}(\vbm) = 
        \begin{pmatrix}
           \sin\left(2\pi \Bbm\vbm\right),\cos\left(2\pi \Bbm\vbm\right)
         \end{pmatrix},\quad\text{with}\quad\vbm=(x,y,z),
\end{align}
where $\Bbm\in\R^{ \times 3}$ is a random matrix with i.i.d Gaussian elements. 
Supplementary Figure~\ref{Fig:encoding_ablation} and~\ref{Fig:Encoding} show how positional encoding gives overly smooth features and Gaussian encoding leads to noise patterns. Radial encoding addresses these two problems by incorporating an extra rotation mapping $\Rbm_\theta$ into the positional encoding formulation 
\begin{align}
\label{Eq:Rad}
\gamma_\text{rad}(\vbm) = 
        \begin{pmatrix}
           \sin\left(2^0\pi\Rbm_{\bm{\theta}}\vbm\right),\cos\left(2^0\pi\Rbm_{\bm{\theta}}\vbm\right), \\
           \vdots \\
           \sin(2^{L-1}\pi\Rbm_{\bm{\theta}}\vbm),\cos(2^{L-1}\pi\Rbm_{\bm{\theta}}\vbm)
         \end{pmatrix} \\
         \quad\text{with}\quad\Rbm_{\bm{\theta}}=
         \left\{\begin{bmatrix}
           \cos(\theta_k)& -\sin(\theta_k)\\
           \sin(\theta_k) & \cos(\theta_k)
         \end{bmatrix}\right\}_{k=1}^K. \nonumber
\end{align}
The rotation mappings enables MLP to respond to features occurring at different orientations without being completely random, leading to better recovery of fine features by avoiding noise. Supplementary Figure~\ref{Fig:encoding_ablation} and~\ref{Fig:Encoding} visually show the benefit of radial encoding. Supplementary Table~\ref{Tab:Encoding}  numerically quantifies the improvement. Please see detailed discussion in \emph{Ablation Experiments}.

\section*{Additional Technical Details} 

\textbf{Deep denoiser in the noise-reduction regularizer.} We adopted DnCNN~\cite{Zhang.etal2017} as our deep denoiser. The network structure is illustrated in Supplementary Figure~\ref{Fig:dncnn}. The network is composed of ten $3\times3$ convolutional layers, of which the first nine layers are activated by rectified linear unit (ReLU) and the last layer is not equipped with any activation function.
Prior to the inclusion into \proposed, the DnCNN network, $\Rsf_\sigma$, is pre-trained to map the noisy input to the \emph{noise residual}. Subsequently, the final $\Dsf_\sigma$ and noise reduction regularizer $\Rcal_\text{NR}$ are expressed as 
\begin{equation}
\Dsf_\sigma \defn \Isf - \Rsf_\sigma\quad\text{and}\quad\Rcal_\text{NR}(\xbm) = \|\Rsf_\sigma(\xbm)\|_2^2.
\end{equation} 
The training dataset for DnCNN is synthesized by adding additive white Gaussian noise (AWGN) of strength $\sigma$ to the natural images from BSD500 dataset~\cite{Martin.etal2001}. Note that the decoupling of the denoiser and the imaging modality has been shown successful in the existing plug-and-play literature~\cite{Sun.etal2020,Xu.etal2020,Zhang.etal2017a}. 
We trained DnCNN by using Adam optimizer for five noise levels $\sigma\in\{1,2,3,4,5\}$ and selected the one that leads to the best visual performance for each sample.
The loss function is defined as
\begin{equation}
\Lcal(\Ibm_\text{input},\Ibm_\text{noise}) = \|\Rsf_\sigma(\Ibm_\text{input})-\Ibm_\text{noise}\|_2^2 + \|\Rsf_\sigma(\Ibm_\text{input})-\Ibm_\text{noise}\|_1
\end{equation}
where $\Ibm_\text{input}$ is the input noisy image, and $\Ibm_\text{noise}$ is the ground-truth noise. 

\vspace{0.5em}
\noindent
\textbf{Computational platform \& training statistics.} We trained all of our deep learning models on a machine equipped with one AMD Threadripper 3960X 24-core CPU and four Nvidia RTX 3090 GPUs. We parallelized the training of \proposed~over two GPUs to accelerate the convergence. 
Under this setup, it approximately takes a day ($\sim$ 20 hours) to train the model for each sample.
We implement a decreasing learning rate, which decays exponentially as the training epoch increases, to smooth the optimization.
Supplementary Figure~\ref{Fig:loss} visually illustrates the training progress of \proposed~for each sample.
The mean absolute error (MAE) between the predicted and real test measurements is plotted against the iteration number. Here, we define the MAE as
\begin{equation}
\text{MAE}(\ybm_\text{pred},\ybm_\rho) = \frac{1}{M}\|\ybm_\text{pred}-\ybm_\rho\|_1,
\end{equation}
where $\ybm_\text{pred}$ denotes the full measurements generated by the entire predicted RI volume.

\section*{Ablation Experiments}
In this section, we present our ablation studies of \proposed focusing on three aspects, that is, (1) comparing different $x$-$y$ encoding strategies, (2) demonstrating the implicit regularization of the MLP network, and (3) illustrating the effect of explicit regularization. 

\vspace{0.5em}
\noindent
\textbf{Comparison of different $x$-$y$ encodings on experimentally collected samples.}
We validated our \emph{Radial} encoding by comparing it against the \emph{Positional}~\cite{Mildenhall.etal2020} and \emph{Gaussian}~\cite{Tancik.etal2020} encodings on C.\ elegans, which contains small biological features that are difficult to recover.
We controlled all the remaining configuration of \proposed~to be the same in order to observe the influence of the encoding strategies.
The encoding of $z$ is set constantly to positional encoding with $L_z=6$.
Supplementary Figure~\ref{Fig:encoding_ablation} presents the experimental results with visual difference highlighted by arrows.
By comparing the axial slices shown in Supplementary Figure~\ref{Fig:encoding_ablation}(a)-\ref{Fig:encoding_ablation}(c), one can observe that Radial leads to a clearer recovery than the other two strategies. 
For example, the body wall and anterior pharyngeal bulb of Radial are sharp and refined, while that of Positional and Gaussian are either oversmoothed or noisy.
The lateral slices shown in Supplementary Figure~\ref{Fig:encoding_ablation}(d)-\ref{Fig:encoding_ablation}(f) confirm our observation.
Although Positional provides a brighter visualization of the grinder, Radial better visualizes other biological features that are highlighted by arrows.
We report that radial encoding leads to similar visual improvement in all samples.

\vspace{0.5em}
\noindent
\textbf{Quantitative evaluation of different $x$-$y$ encodings.} We quantitatively evaluated the performance of the \emph{Positional}, \emph{Gaussian}, and \emph{Radial (Ours)} encodings.  We adopted the 3D cell phantom used by Yanny. \emph{et al}~\cite{Yanny.etal2022} for the simulation. We denoised the original sample using the average filtering followed by clipping. 
The phantom contains $486\times486\times32$ voxels, with each has the size of $0.1625\times0.1625\times0.1\,\mu\text{m}^3$.
The maximal value of RI is set to $\nbm_\text{re} = 0.075$ and $\nbm_\text{im} = 0$, meaning that there is no absorption.
The simulation is based on the aIDT setup and used the \emph{split-step non-paraxial (SSNP)} simulator to simulate the full wave propagation~\cite{Lim.etal2019}. Recall that the configuration includes an annular LED array for illumination and an objective that has the numerical aperture (NA) equal to $0.65$.
During the acquisition, the phantom is assumed to be immersed in the vacuum ($\nbm_0=1$), and a total of 24 measurements are acquired using the light of the wavelength equal to $515$ nm.
Supplementary Table~\ref{Tab:Encoding} summarizes the PSNR and MSE values achieved by different encodings.
We note that Radial achieves better results than Positional and Gaussian. Intuitively, this is because Radial encoding enables more structural flexibility by incorporating the rotation mappings (see detailed explanation in \emph{Radial Encoding}).
Supplementary Figure~\ref{Fig:Encoding} provides visual comparison on two axial slices. In the zoom-in region, Positional and Gaussian lead to either overly smooth or noisy reconstructions, while Radial leads to the best results. 

\vspace{0.5em}
\noindent
\textbf{Demonstration of implicit regularization on experimentally collected samples.} We demonstrate the implicit regularization of the MLP network on spirogyra algae.
In the experiment, we reconstructed the sample by using \proposed~equipped with explicit regularization (\emph{\proposed}) and without explicit regularization (\emph{\proposed-Noreg}).
We control the remaining configuration of \proposed~to be the same for fair comparison. 
We additionally included SIMBA and Tikhonov as baseline methods.
Supplementary Figure~\ref{Fig:regularization_ablation_algae} presents the results with visual differences highlighted by boxes. 
By comparing Supplementary Figure~\ref{Fig:regularization_ablation_algae}(b), \ref{Fig:regularization_ablation_algae}(c), and \ref{Fig:regularization_ablation_algae}(d), one can observe that \proposed-Noreg visually outperforms SIMBA and Tikhonov by a large margin, clearly showing that MLP by itself offers implicit regularization.
One can additionally observe that \proposed-Noreg nearly matches the performance of the full \proposed~by comparing Supplementary Figure~\ref{Fig:regularization_ablation_algae}(a) and \ref{Fig:regularization_ablation_algae}(b), though \proposed~better alleviates the dark-shade artifacts highlighted by the boxes.

\vspace{0.5em}
\noindent
\textbf{Demonstration of explicit regularization on experimentally collected samples.}
We now show that explicit regularization enables \proposed~to address more challenging samples that suffer from scattering-related artifacts. We consider C.\ elegans (body) and four \proposed~models: full regularization (\emph{\proposed}), only axial-continuity regularization (\emph{\proposed-AC}), only noise-reduction regularization (\emph{\proposed-NR}), and no explicit regularization (\emph{\proposed-Noreg}).
We set all the remaining parameters to be the same. Supplementary Figure~\ref{Fig:regularization_ablation} presents the results with visual differences highlighted by arrows. Note how \proposed-Noreg fails to reconstruct the worm (Supplementary Figure~\ref{Fig:regularization_ablation}(d)) when the external regularization is removed, indicating that the implicit regularization posed by MLP is insufficient for correcting for artifacts in thick/strongly-scattering objects.
By including $2$D noise reduction (Supplementary Figure~\ref{Fig:regularization_ablation}(c)), \proposed-NR significantly improves the imagery quality over \proposed-Noreg, but still leads to axial inconsistency as shown in the lateral view (Supplementary Figure~\ref{Fig:regularization_ablation}(g)).
By changing the regularization to axial continuity (Supplementary Figure~\ref{Fig:regularization_ablation}(b)), \proposed-AC improves both axial and lateral results, but causes the cross-slice interference as shown in the expanded regions highlighted by arrows.
On the other hand, \proposed~achieves the best reconstruction performance by synergistically leveraging both regularization strategies (Supplementary Figure~\ref{Fig:regularization_ablation}(a)). 
Supplementary Figure~\ref{Fig:regularization_ablation}(a) and \ref{Fig:regularization_ablation}(d) clearly show the necessity of explicit regularization under stronger scattering.
Visual comparison of lateral slices in Supplementary Figure~\ref{Fig:regularization_ablation}(e)-Supplementary Figure~\ref{Fig:regularization_ablation}(h) demonstrates the same evolution.

\vspace{0.5em}
\noindent
\textbf{Quantitative evaluation of different regularizers.} We quantitatively compared the influence of regularizers on the simulated granulocyte cell cluster. 
We refer to the \emph{Quantitative Evaluation} in the main paper for the experimental setup. 
Supplementary Table~\ref{Tab:NumericalComparison} summarizes the  PSNR and MSE values achieved by every \proposed~variant and the two baseline methods.
We note that both SIMBA and Tikhonov are based on discrete voxel-grid volume representation.   
By comparing \proposed-Noreg and SIMBA, it is clear that the MLP representation itself provides regularization. For example, \proposed-Noreg achieves the PSNR value of $23.69$ dB, enabling $1.3$ dB and $2.97$ dB improvement over SIMBA and Tikhonov, respectively. 
On the other hand, the best results are achieved by additionally including explicit regularizers. The results of \proposed-NR and \proposed-AC show that the exclusion of either noise reduction and axial continuity will cause at least $0.16$ dB degradation of the quality. The synergistic combination of the internal and external regularization leads to improvements of 1.64 dB and 3.31 dB in PSNR (corresponding to $1.5\times$ and $2.1\times$ MSE reduction).

\newpage
\setcounter{figure}{0}
\renewcommand{\figurename}{Supplementary Figure}
\renewcommand{\tablename}{Supplementary Table}

\begin{figure}[t!]
\begin{center}
\includegraphics[width=\linewidth]{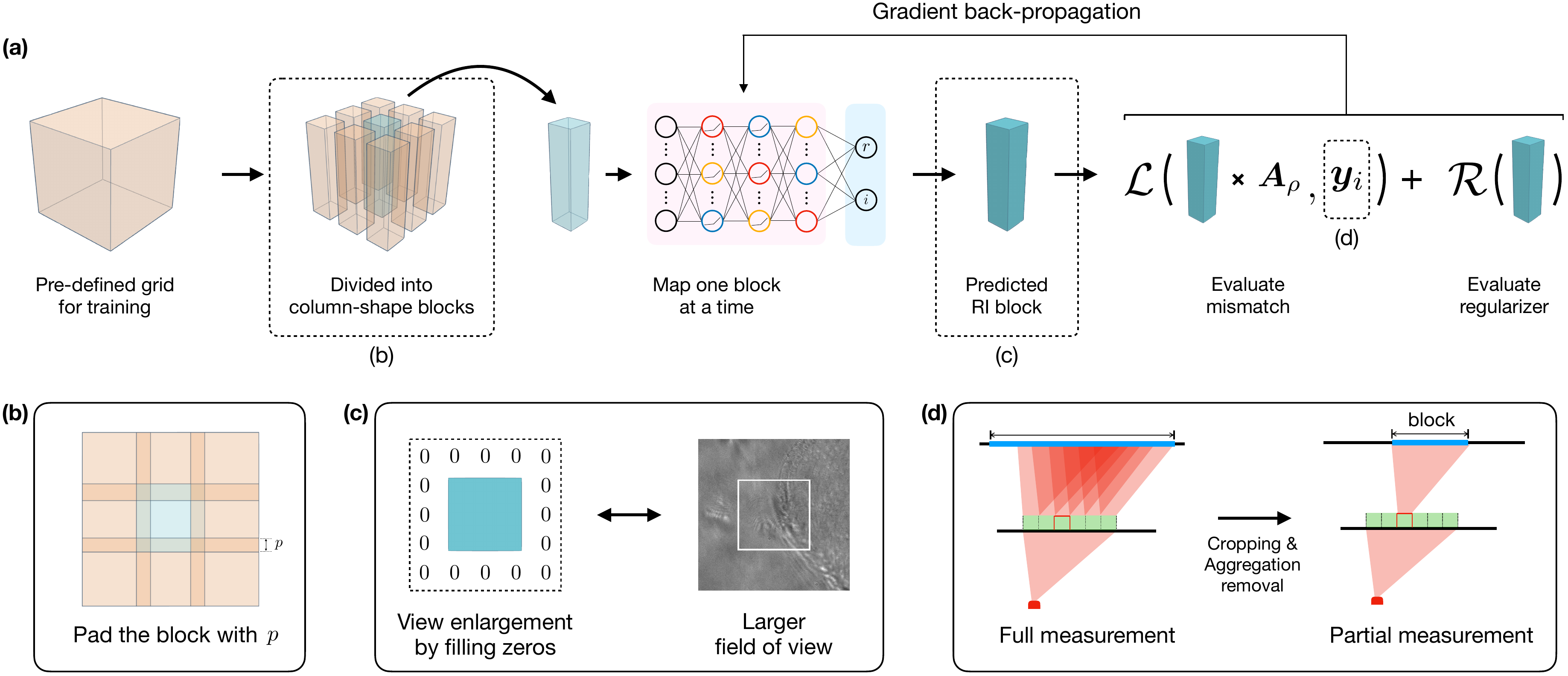}
\end{center}
\caption{
\textbf{Visual illustration of the block-wise training procedure for \proposed.}
\textbf{(a)} General workflow of the proposed training procedure.
\textbf{(b)} Illustration of \emph{padding}.
\textbf{(c)} Illustration of \emph{view enlargement}.
\textbf{(c)} Illustration of \emph{measurement separation}.
}
\label{Fig:training}
\end{figure}
\clearpage

\begin{figure}[t!]
\begin{center}
\includegraphics[width=0.85\linewidth]{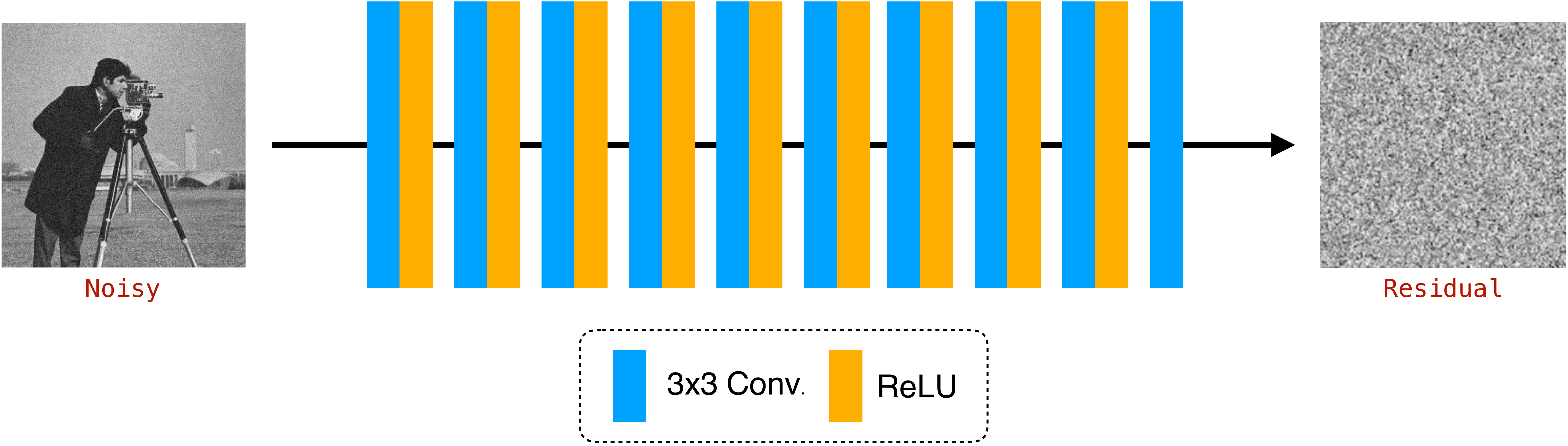}
\end{center}
\caption{
\textbf{Network architecture of DnCNN.} DnCNN is trained to map the noisy input to the noise residual by using the BSD500 training dataset~\cite{Martin.etal2001}, which consists only of natural images. 
}
\label{Fig:dncnn}
\end{figure}
\clearpage

\begin{figure}[t!]
\begin{center}
\includegraphics[width=0.85\linewidth]{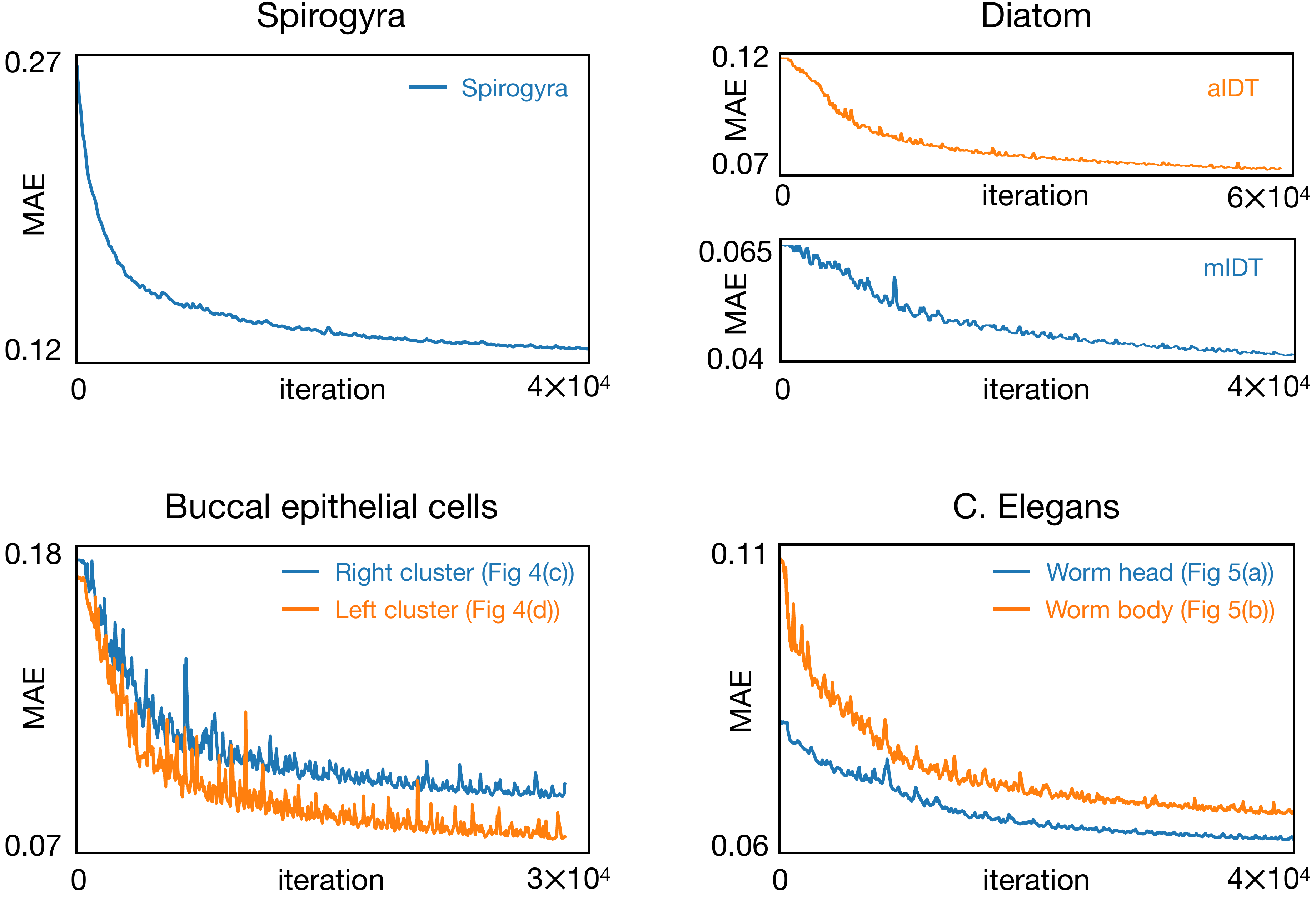}
\end{center}
\caption{
\textbf{Convergence of \proposed~for different real biological samples.}
In each figure, the mean absolute error (MAE) between the predicted and real test measurements is plotted against the iteration number.
}
\label{Fig:loss}
\end{figure}

\begin{figure}[t!]
\begin{center}
\includegraphics[width=\linewidth]{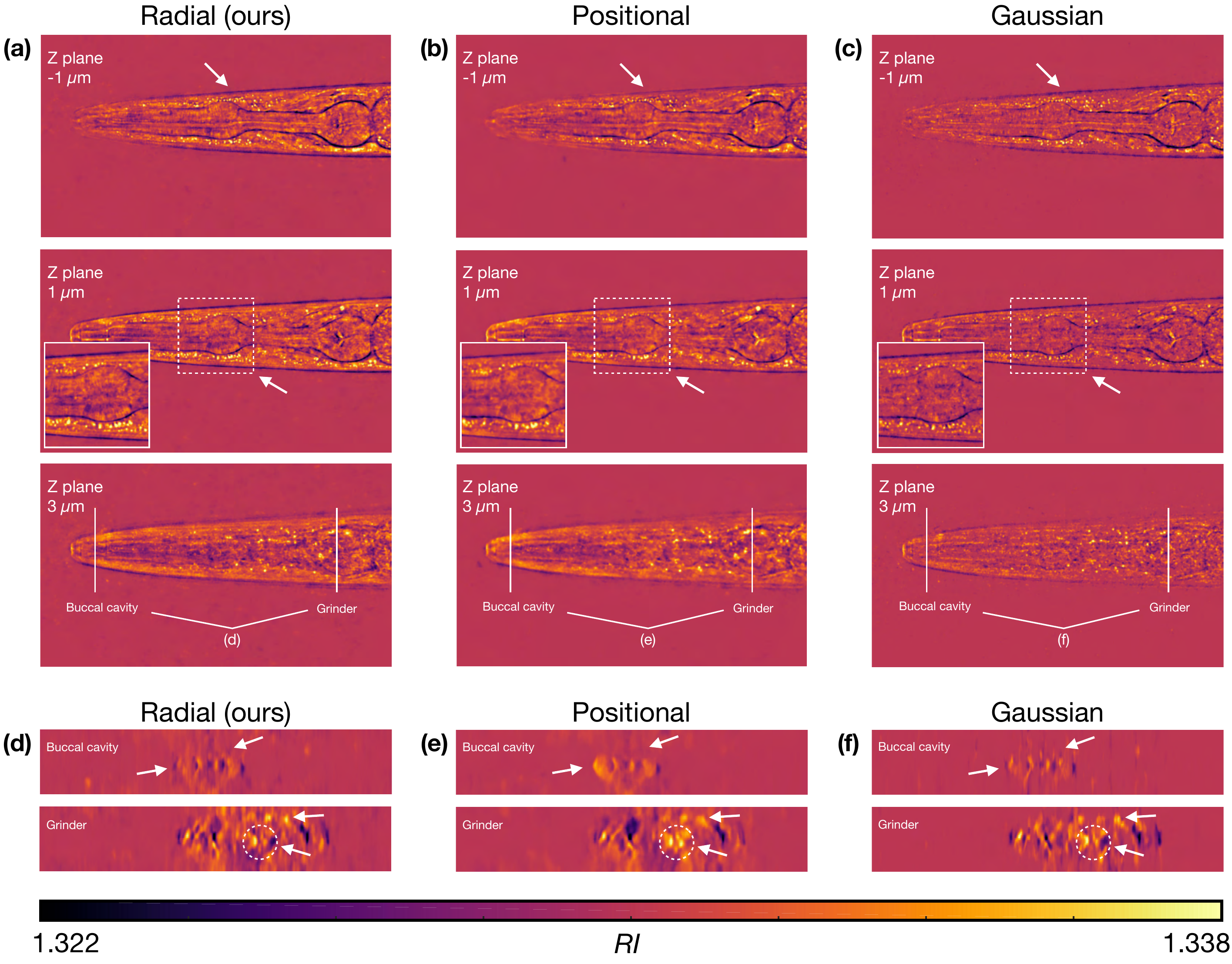}
\end{center}
\caption{
\textbf{Visual comparison of different $x$-$y$ encoding strategies on \emph{C.\ Elegans (head)}. Visual differences are highlighted using arrows.}
\textbf{(a), (b) \& (c)} Axial slices of the C. elegans' head at $z\in\{-1,1,3\}\,\mu$m reconstructed by using \emph{Radial (ours)}, \emph{Positional}, and \emph{Gaussian} encoding of the $x$-$y$ plane. 
The encoding of $z$, as well as the rest of \proposed, is set to be the same in the comparison.
\textbf{(d), (e) \& (f)} The associated lateral slices of the buccal cavity and grinder shown in the axial views.}
\label{Fig:encoding_ablation}
\end{figure}
\clearpage

\begin{figure}[t]
\begin{center}
\includegraphics[width=\linewidth]{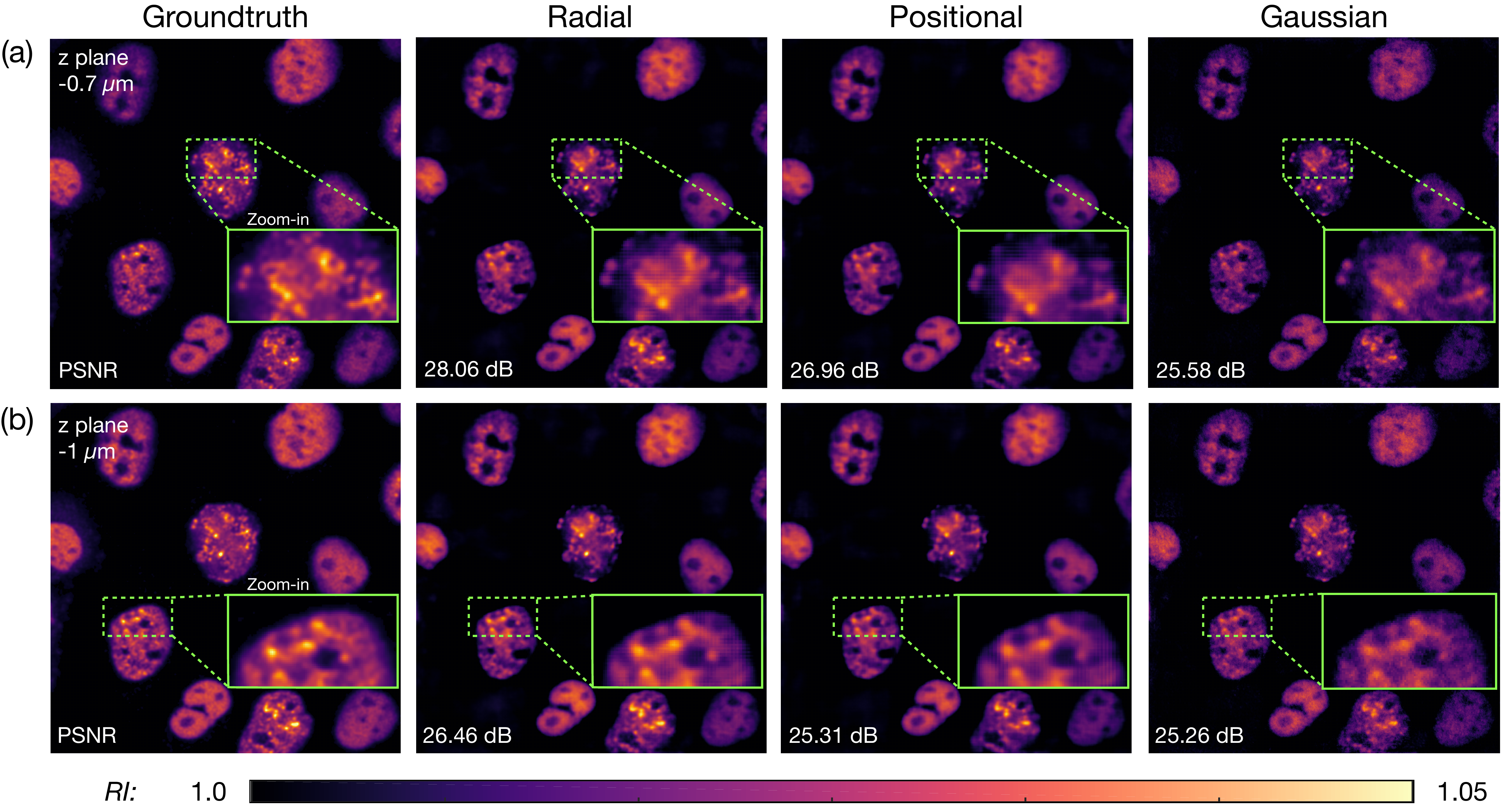}
\end{center}
\caption{
\textbf{Performance comparison of \emph{Positional}~\cite{Mildenhall.etal2020}, \emph{Gaussian}~\cite{Tancik.etal2020}, and \emph{Radial (ours)} encodings on the \emph{Yanny Sample}~\cite{Yanny.etal2022}.} \textbf{(a)} $x$-$y$ slice at the depth $z=-0.7\,\mu$m. \textbf{(b)} $x$-$y$ slice at the depth $z=-1.0\,\mu$m. Visual differences are enlarged by the boxes.}
\label{Fig:Encoding}
\end{figure}
\clearpage

\begin{figure}[t!]
\begin{center}
\includegraphics[width=\linewidth]{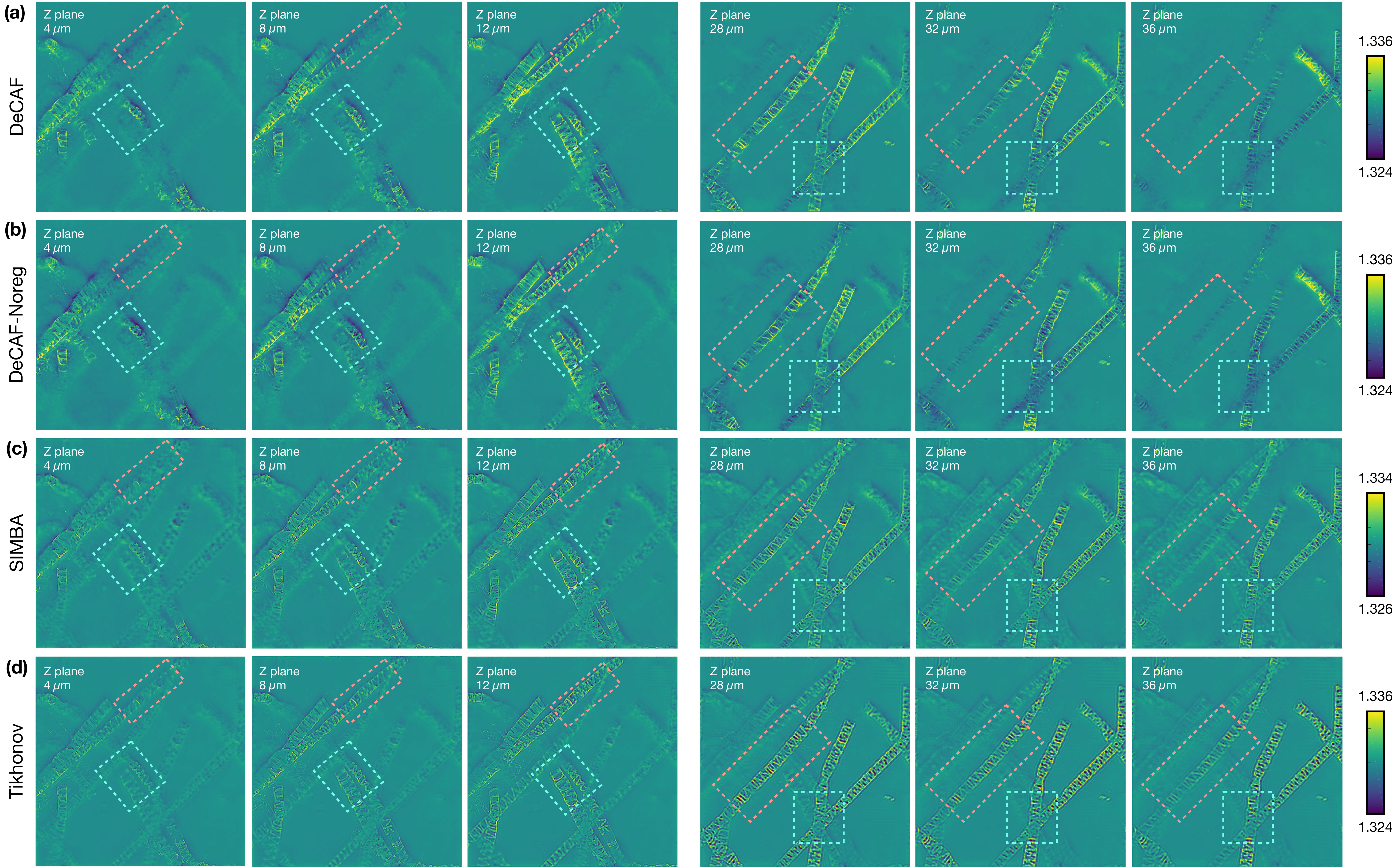}
\end{center}
\caption{
\textbf{Demonstration of the \emph{implicit regularization} due to the MLP on \emph{Spirogyra Algae}. Visual differences are highlighted using boxes.}
\textbf{(a)} \proposed, which includes an explicit regularizer.
\textbf{(b)} \proposed-Noreg relies only on implicit regularization by MLP.
\textbf{(c)} SIMBA.
\textbf{(d)} Tikhonov.
Axial slices at $\{4,8,12\}\mu$m and $\{28,32,36\}\mu$m are selected. 
}
\label{Fig:regularization_ablation_algae}
\end{figure}
\clearpage

\begin{figure}[t!]
\begin{center}
\includegraphics[width=\linewidth]{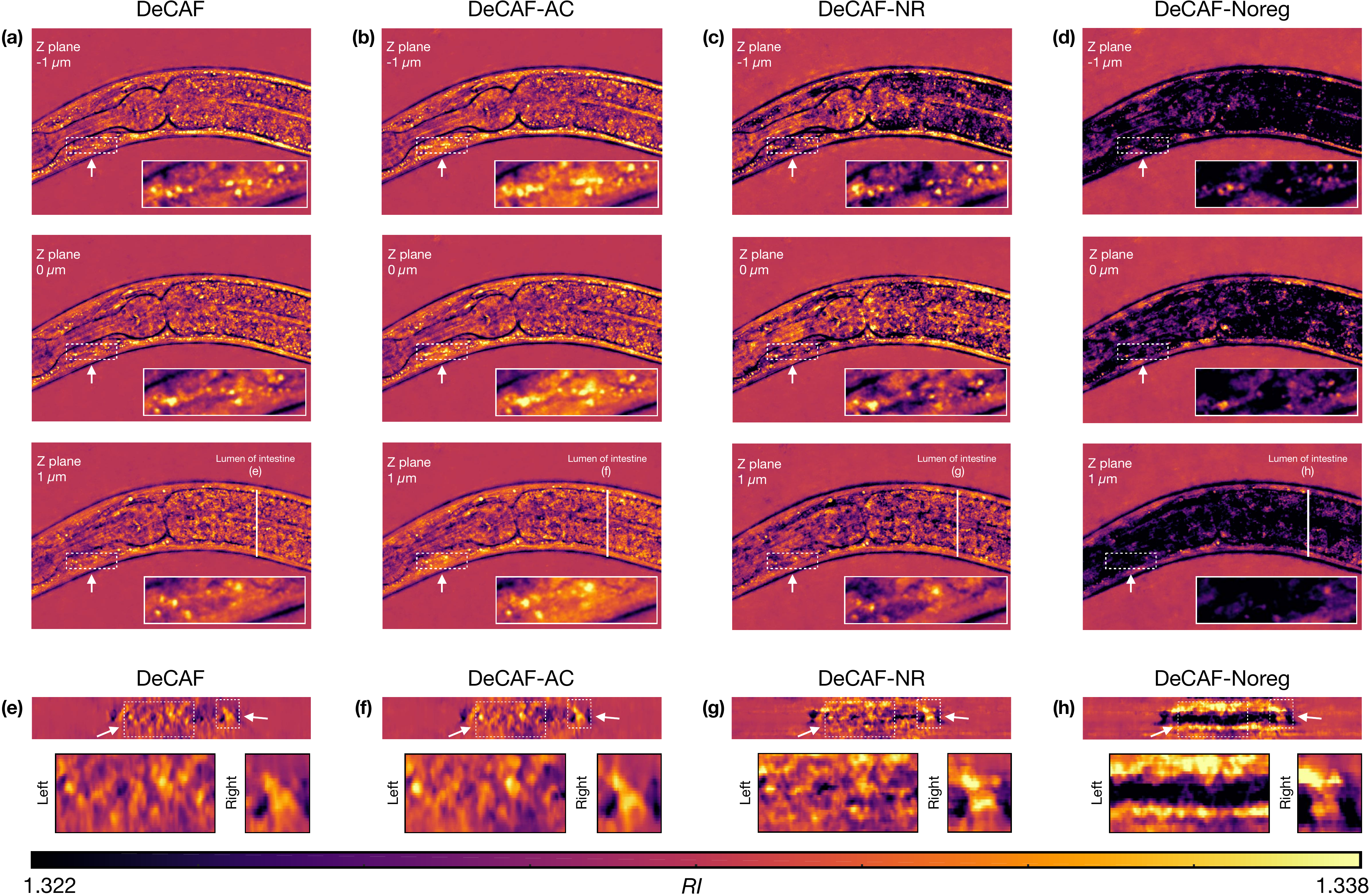}
\end{center}
\caption{
\textbf{Demonstration of the effectiveness of explicit regularization on \emph{C.\ Elegans (body)}. Visual differences are highlighted by arrows.}
\textbf{(a), (b) \& (c)} Axial slices of the C. Elegans body at $z\in\{-1,1,3\}\,\mu$m reconstructed by using full regularization (\emph{\proposed}), only the axial continuity (\emph{\proposed-AC}), only the noise reduction (\emph{\proposed-NR}), and no regularization (\emph{\proposed-Noreg}). 
The rest of \proposed~is set to be same in the comparison.
\textbf{(d), (e) \& (f)} The associated lateral slices of the lumens of intestine shown in the axial views.}
\label{Fig:regularization_ablation}
\end{figure}
\clearpage


\definecolor{cerulean}{rgb}{0.0, 0.48, 0.65}
\definecolor{cadmiumred}{rgb}{0.89, 0.0, 0.13}

\begin{table*}[t]
	\centering
	\caption{
	Comparison of the PSNR and MSE values achieved by \emph{Positional}~\cite{Mildenhall.etal2020}, \emph{Gaussian}~\cite{Tancik.etal2020}, and \emph{Radial (ours)} encodings on the \emph{Yanny Sample}~\cite{Yanny.etal2022}. Note that all metrics are evaluated for the entire volume, and the best values are highlighted in \emph{bold}.}
	\label{Tab:Encoding}
	\small
	\begin{tabular*}{265pt}{R{70pt}C{1pt}C{65pt}C{65pt}C{0pt}} 	
		\toprule
		\textbf{Encoding} & & \textbf{PSNR} $\color{cadmiumred}\uparrow$ & \textbf{MSE} $\color{cerulean}\downarrow$\\
		\cmidrule[0.5pt]{1-1}\cmidrule[0.5pt]{3-5}
		Positional        &  & $25.88$ dB & $1.18\times10^{-5}$ &\\ [0.7ex]
		Gaussian        &  & $25.96$ dB & $1.16\times10^{-5}$ &\\ [0.7ex]
		\hdashline\noalign{\vskip 1.3mm}
		Radial (Ours)              &  & \textbf{$\mathbf{26.09}$ dB} & \textbf{$\mathbf{1.12\times10^{-5}}$} & \\ [0.3ex]
		\bottomrule
	\end{tabular*}
\end{table*}
\clearpage

\definecolor{cerulean}{rgb}{0.0, 0.48, 0.65}
\definecolor{cadmiumred}{rgb}{0.89, 0.0, 0.13}
\begin{table*}[t]
	\centering
	\caption{
	Numerical values achieved by Tikhonov, SIMBA, and \proposed~on the simulated granulocyte cell phantom. Results achieved by different \proposed~variants are also included. Best values are highlighted in \emph{bold}.}
	\label{Tab:NumericalComparison}
	\small
	\begin{tabular*}{260pt}{R{70pt}C{1pt}C{65pt}C{65pt}C{0pt}} 	
		\toprule
		\textbf{Methods} & & \textbf{PSNR} $\color{cadmiumred}\uparrow$ & \textbf{MSE} $\color{cerulean}\downarrow$\\
		\cmidrule[0.7pt]{1-1}\cmidrule[0.5pt]{3-5}
		Tikhonov~\cite{Ling.etal18}     &   & $20.72$ dB & $4.77\times10^{-5}$ &\\ [0.3ex]
		SIMBA~\cite{Wu.etal2020}      &   & $22.39$ dB & $3.24\times10^{-5}$ &\\ [1.2ex]
		\hdashline\noalign{\vskip 1.3mm}
		\proposed-Noreg   &   & $23.69$ dB & $2.41\times10^{-5}$ &\\ [0.7ex]
		\proposed-NR        &   & $23.75$ dB & $2.37\times10^{-5}$ &\\ [0.7ex]
		\proposed-AC        &   & $23.87$ dB & $2.31\times10^{-5}$ &\\ [0.7ex]
		\hdashline\noalign{\vskip 1.3mm}
		\proposed              &   & \textbf{$\mathbf{24.03}$ dB} & \textbf{$\mathbf{2.22\times10^{-5}}$} & \\ [0.3ex]
		\bottomrule
	\end{tabular*}
\end{table*}
\clearpage

\end{document}